\documentclass[%
 aip,
 amsmath,amssymb,
preprint,%
]{revtex4-2}

\usepackage{graphicx}
\usepackage{dcolumn}
\usepackage{bm}

\usepackage[utf8]{inputenc}
\usepackage[T1]{fontenc}

\usepackage{braket} 
\usepackage{comment} 
\usepackage{colortbl}
\definecolor{Gray}{gray}{0.9}

\newcommand{\be}{\begin{equation}}
\newcommand{\ee}{\end{equation}}
\newcommand{\ti}[1]{\text{#1}}
\newcommand{\mc}[1]{\mathcal{#1}}
\newcommand{\hmc}[1]{\hat{\mathcal{#1}}}

\newcommand{\fig}[1]{Fig.~\ref{#1}}

\newcommand{\Fig}[1]{Figure~\ref{#1}}

\newcommand{\eq}[1]{Eq.~\eqref{#1}}
\newcommand{\eqs}[2]{Eqs.~\eqref{#1} and \eqref{#2}}
\newcommand{\Eq}[1]{Equation~\eqref{#1}}

\newcommand{\stn}[1]{Sec.~\ref{#1}}
\newcommand{\Stn}[1]{Section~\ref{#1}}
\newcommand{\app}[1]{Appendix~\ref{#1}}

\newcommand{\bea}{\begin{eqnarray}}
\newcommand{\eea}{\end{eqnarray}}
 
\newcommand{\ba}{\begin{array}}
\newcommand{\ea}{\end{array}}

\newcommand{\bl}{\begin{flalign}}
\newcommand{\enl}{\end{flalign}}

\newcommand{\tr}{\text{Tr}}

\usepackage[normalem]{ulem} 
\newcommand{\mkadd}[1]{{\color{red}{#1}}}
\newcommand{\mkerase}[1]{\ifmmode{\color{red}{\text{\sout{\ensuremath{#1}}}}}\else{\color{red}{\sout{#1}}}\fi}

\begin{document}

\preprint{AIP/123-QED}

\title{General Framework for Quantifying Dissipation Pathways in Open Quantum Systems. II. Numerical Validation and the Role of Non-Markovianity}

\author{Chang Woo Kim}
\affiliation{
    Department of Chemistry, Chonnam National University, Gwangju 61186, South Korea
    }
\email{cwkim66@jnu.ac.kr}
    
\author{Ignacio Franco}
\affiliation{
    Department of Chemistry, University of Rochester, Rochester, New York 14627, USA
    }
\affiliation{
    Department of Physics, University of Rochester, Rochester, New York 14627, USA
    }
\email{ignacio.franco@rochester.edu}

\date{\today}

\begin{abstract}
In the previous paper [C. W. Kim and I. Franco, J.~Chem.~Phys.~\textbf{160},~214111~(2024)], we developed a theory called MQME-D, which allows us to decompose the overall energy dissipation process in open quantum system dynamics into contributions by individual components of the bath when the subsystem dynamics is governed by a Markovian quantum master equation (MQME). Here, we contrast the predictions of MQME-D against the numerically exact results obtained by combining hierarchical equations of motion (HEOM) with a recently reported protocol for monitoring the statistics of the bath. Overall, MQME-D accurately captures the contributions of specific bath components to the overall dissipation while greatly reducing the computational cost as compared to exact computations using HEOM. The computations show that MQME-D exhibits errors originating from its inherent Markov approximation. We demonstrate that its accuracy can be significantly increased by incorporating non-Markovianity by exploiting time scale separations (TSS) in different components of the bath. Our work demonstrates that MQME-D combined with TSS can be reliably used to understanding how energy is dissipated in realistic open quantum system dynamics.
\end{abstract}

\maketitle

\section{Introduction}\label{section:introduction}

In Paper I\cite{Kim2024} of our two-paper series, we introduced a new theoretical framework that can be used to decompose the energy dissipated in the open quantum system dynamics into contributions from individual components of the bath. The new framework, referred to as MQME-D, targets the Markovian quantum master equations derived from the Nakajima-Zwanzig projection operator technique.\cite{Nakajima1958,Zwanzig1960} By using it, we recovered the formulae for quantifying dissipation pathways reported in our earlier work,\cite{Kim2021} which were developed in the context of F{\"o}rster resonance energy transfer (FRET). We also derived and presented the expressions for other types of subsystem-bath interactions such as linearly coupled spin environment. In addition, the framework also allowed us to construct rigorous analytical proofs on thermodynamic principles such as energy conservation and detailed balance.

Here, we investigate the accuracy and computational efficiency of MQME-D as a method to decompose the overall dissipation during the open quantum system dynamics. An initial assessment of the strategy was done in Ref.~\citenum{Kim2021} based on a comparison with a mixed-quantum classical (MQC) method.\cite{Kim2020} However, the accuracy of the MQC method was limited by the zero-point energy leak and partial neglect of the interplay between the subsystem and bath,\cite{Nassimi2010,Kim2014} which both originate from the classical treatment of the trajectories.

Such an experience led us to develop alternative methods that do not rely on classical description of the bath. As a result, we successfully constructed a computational method\cite{Kim2022} to quantify dissipation by individual bath modes based on numerically exact simulation methods such as hierarchical equations of motion (HEOM)\cite{Tanimura1990,Tanimura2020} or quasi-adiabatic propagator path integral (QUAPI).\cite{Topaler1993,Kundu2023} Using this method, we now can systematically assess the reliability of MQME-D by employing a numerically exact benchmark.


In this paper, we thoroughly examine the performance of MQME-D in various Hamiltonian models and simulation conditions with different subsystem-bath coupling strengths and temperatures. We find that, even when the subsystem dynamics is quite accurate, the dissipation predicted by MQME-D exhibits some disagreements with the benchmark calculation. Careful analysis reveals that a significant portion of the error is caused by the Markov approximation behind MQME, which does not properly reflect the response of the bath during the dynamics. We then incorporate non-Markovianity in the simulation by combining MQME-D with the time scale separation (TSS) method,\cite{MontoyaCastillo2015} and demonstrate that the combination can substantially enhance the accuracy of the decomposition of the dissipation in some cases. In TSS, one separates the bath modes into slow and fast components depending on their characteristic frequencies. Only the fast component directly participates in the dynamics, while the effect of the slow component manifests as the static disorder which introduces the non-Markovianity. In the end, MQME-D with TSS (MQME-D+TSS) achieves nearly quantitative resolution of the dissipated energy in the frequency domain, demonstrating its capability to elucidate the key DOFs that govern the dynamics of realistic systems such as photosynthetic complexes or extended molecular aggregates.

The structure of the paper is as follows. \Stn{section:background} provides a brief overview of MQME-D and the Hamiltonian models used in this work. \Stn{section:accuracy} applies the bare MQME-D to the Hamiltonian of a molecular dimer and contrasts the results against a numerically exact method. \Stn{section:nonMarkovian} introduces the TSS strategy and benchmarks the accuracy of MQME-D+TSS by using different types of Hamiltonian models such as molecular dimer and spin-boson model. \Stn{section:efficiency} presents a brief comparison between the computational efficiencies of MQME-D+TSS and numerically exact methods. Finally, \stn{section:Conclusion} concludes the paper by summarizing the main findings and discussing conceivable research directions for the future.

\section{Theoretical Background}\label{section:background}
The MQME-D theory was developed and discussed in Paper I\cite{Kim2024}. For convenience, below we summarize the main ideas as needed to introduce the simulation strategy and subsequent discussion. First, \stn{subsection:Hamiltonian} introduces the Hamiltonian models used in this work. Then, \stn{subsection:equations} presents the key equations of MQME-D for simulating the dissipation in the models presented in \stn{subsection:Hamiltonian}.

\subsection{Model Hamiltonian}\label{subsection:Hamiltonian}
To test the numerical accuracy of the theory, we study the dissipation in prototypical models of open quantum systems involving harmonic bath modes. However, as discussed in Ref.~\citenum{Kim2024}, the MQME-D framework is applicable to any (harmonic or anharmonic) environment as long as it is composed of independent bath degrees of freedom. The Hamiltonian of an open quantum system is written as
\begin{equation}\label{eq:H}
    \hat{H} = \hat{H}_\ti{sub} + \hat{H}_\ti{bath} + \hat{H}_\ti{int},
\end{equation}
where $\hat{H}_\ti{sub}$, $\hat{H}_\ti{bath}$, and $\hat{H}_\ti{int}$ are the Hamiltonian components describing the subsystem, the bath, and the subsystem-bath interaction, respectively. By denoting the subsystem states as $\{\ket{A}\}$, $\hat{H}_\ti{sub}$ is written as
\begin{equation}\label{eq:H_sub}
    \hat{H}_\ti{sub} = \sum_A E_A \ket{A}\bra{A} + \sum_A \sum_{B<A} V_{AB} (\ket{A}\bra{B} + \ket{B}\bra{A}),
\end{equation}
where $E_A$ is the energy of the subsystem state $\ket{A}$ and $V_{AB}$ is the coupling between the subsystem states $\ket{A}$ and $\ket{B}$.
The harmonic bath modes and their interaction with the subsystem are described by
\begin{equation}\label{eq:H_bath}
    \hat{H}_\ti{bath} = \sum_j \bigg( \frac{\hat{p}_j^2}{2} + \frac{\omega_j^2 \hat{x}_j^2}{2} \bigg),
\end{equation}
\begin{equation}\label{eq:H_int}
    \hat{H}_\ti{int} = \sum_A \bigg[ \ket{A} \bra{A} \otimes \sum_j \bigg( - \omega_j^2 d_{Aj} \hat{x}_j + \frac{\omega_j^2 d_{Aj}^2}{2} \bigg) \bigg].
\end{equation}
In the above, $\omega_j$ is the characteristic frequency of the $j$-th bath mode, whose momentum and position are described by operators $\hat{p}_j$ and $\hat{x}_j$, respectively. The state-dependent parameter $d_{Aj}$ quantifies the strength of the subsystem-bath coupling and corresponds to the location of the minimum in the potential energy surface (PES) $\hat{h}_A = \bra{A} \hat{H} \ket{A}$ along the coordinate $x_j$. Overall, \eq{eq:H_int} assumes that the bath only couples to diagonal part of $\hat{H}_\ti{sub}$ (Condon approximation) with linear dependence in $\{\hat{x}_j\}$. The strength of the subsystem-bath interaction is collectively described by the bath spectral density (BSD), which is defined as
\begin{equation}\label{eq:BSD}
    J_{AB}(\omega) = \sum_j \frac{\omega_j^3 d_{Aj} d_{Bj}}{2} \delta(\omega - \omega_j),
\end{equation}
which is related to the generalized reorganization energy $\Lambda_{AB}$ via
\begin{equation}\label{eq:spd_reorg_relation}
    \Lambda_{AB} = \sum_j \frac{\omega_j^2 d_{Aj} d_{Bj}}{2} = \int_0^\infty \frac{J_{AB} (\omega)}{\omega} \; d\omega.
\end{equation}
Analytical expressions of the BSD are frequently used as approximate models for describing realistic systems. One widely used form of the BSD is the Drude-Lorentz distribution expressed as
\begin{equation}\label{eq:DL_BSD}
    J_\ti{DL}(\omega) = \frac{2 \Lambda}{\pi} \frac{\omega_\ti{c} \omega}{\omega^2 + \omega_\ti{c}^2},
\end{equation}
which is often used to model the slow relaxation of the solvent due to the exponential form of the corresponding bath time correlation function in the high-temperature limit.\cite{Ikeda2020} In \eq{eq:DL_BSD}, $\Lambda$ is the total reorganization energy of the bath and $\omega_\ti{c}$ is the cutoff frequency which determines the relaxation time of the bath. Another model that will be used in this work is the Brownian oscillator whose BSD is expressed as
\begin{equation}\label{eq:BO_BSD}
    J_\ti{BO}(\omega) = \frac{2 \Lambda \gamma}{\pi} \frac{2 \omega_0^2 \omega}{(\omega^2 - \omega_0^2)^2 + 4\gamma^2 \omega^2}.
\end{equation}
Here, $\omega_0$ is the characteristic frequency of the oscillator and $\gamma$ is the strength of the damping. The time correlation function associated with \eq{eq:BO_BSD} resembles the behavior of a damped harmonic oscillator,\cite{Ikeda2020} which makes $J_\ti{BO}(\omega)$ a realistic model for describing molecular vibrations and photonic cavities.

\subsection{MQME-D for a Bath of Harmonic Oscillators}\label{subsection:equations}

According to Eq.~(15) of Paper I,\cite{Kim2024} the populations of the subsystem states in MQME are governed by a set of coupled first-order rate equations
\begin{equation}\label{eq:FRET_pop}
    \dot{P}_A(t) = \sum_{B \neq A} [-K_{BA} P_A(t) + K_{AB} P_B(t)],
\end{equation}
where $P_A(t)$ is the population of the state $\ket{A}$, and $K_{BA}$ is the rate constant for the population transfer from $\ket{A}$ to $\ket{B}$ whose explicit expression is
\begin{equation}\label{eq:FRET_elec}
    \begin{split}
    K_{BA} &= \frac{2 |V_{AB}|^2}{\hbar^2} \ti{Re} \int_0^\infty \exp \bigg( - \frac{it' (E_B - E_A + \Lambda_{AA} - 2 \Lambda_{AB} + \Lambda_{BB})}{\hbar} \bigg) \\
    &\times \exp [- g_{AA}(t') + 2 g_{AB}(t') - g_{BB}(t') ] \: dt'.
    \end{split}
\end{equation}
Here, $g_{AB}(t')$ is the line broadening function defined as
\begin{equation}\label{eq:lbf}
    g_{AB}(t') = \frac{1}{\hbar} \int_0^\infty J_{AB}(\omega) \bigg[ \coth \bigg( \frac{\beta \hbar \omega}{2} \bigg) \frac{1 - \cos(\omega t')}{\omega^2} + i \: \frac{\sin(\omega t') - \omega t'}{\omega^2} \bigg] \: d\omega,
\end{equation}
where $\beta = 1 / k_\ti{B}T$ is the inverse temperature. With the state populations propagated according to \eq{eq:FRET_elec}, the rate of dissipation $\dot{E}_{j} (t)$ into the $j$-th bath mode can be evaluated by
\begin{equation}
    \dot{E}_{j} (t) = \sum_A \sum_{B < A} [ \mc{K}_{BA}^j P_A(t) + \mc{K}_{AB}^j P_B (t) ],
\end{equation}
where the dissipation rate constants $\{\mc{K}_{BA}^j\}$ are calculated as
\begin{equation}\label{eq:diss_mode}
    \begin{split}
    \mc{K}_{BA}^j &= \frac{2 |V_{AB}|^2}{\hbar^2} (\lambda_{AA}^j - 2 \lambda_{AB}^j + \lambda_{BB}^j) \\
    &\times \ti{Re} \int_0^\infty \exp \bigg( - \frac{it' (E_B - E_A + \Lambda_{AA} - 2 \Lambda_{AB} + \Lambda_{BB})}{\hbar} \bigg) \\
    &\times \exp [- g_{AA}(t') + 2 g_{AB}(t') - g_{BB}(t') ] \\
    &\times \bigg[ \cos(\omega_j t') - i\coth \bigg( \frac{\beta \hbar \omega_j}{2} \bigg) \sin(\omega_j t') \bigg] \: dt'.
    \end{split}
\end{equation}
In the above, $\lambda_{AB}^j = \omega_j^2 d_{Aj} d_{Bj}/2$ is the reorganization energy associated with the $j$-th bath mode. At this point, we introduce specific Hamiltonian models which will be employed in the simulations in \stn{section:accuracy} and \stn{section:nonMarkovian}.

\subsubsection{Local Bath Model}\label{subsubsection:lbm}
In this model, each bath mode only couples to a single subsystem state, and the Hamiltonian can be derived from Eqs.~(\ref{eq:H})--(\ref{eq:H_int}) by setting $d_{Aj} d_{Bj} = 0$ when $A \neq B$. As a result, $\hat{H}_\ti{bath}$ and $\hat{H}_\ti{int}$ are reduced to
\begin{equation}\label{eq:H_bath_loc}
    \hat{H}_\ti{bath} = \sum_A \sum_{j \in A} \bigg( \frac{\hat{p}_j^2}{2} + \frac{\omega_j^2 \hat{x}_j^2}{2} \bigg),
\end{equation}
\begin{equation}\label{eq:H_int_loc}
    \hat{H}_\ti{int} = \sum_A \bigg[ \ket{A} \bra{A} \otimes \sum_{j \in A} \bigg( - \omega_j^2 d_{Aj} \hat{x}_j + \frac{\omega_j^2 d_{Aj}^2}{2} \bigg) \bigg],
\end{equation}
where we have introduced the notation $j \in A$ to express that the $j$-th bath mode exclusively belongs to $\ket{A}$. For this model, the frequency-resolved rate of dissipation can be calculated by Eq.~(75) of Paper I\cite{Kim2024},
\begin{equation}\label{eq:diss_dens}
    \mc{D}_A(\omega, t) = \sum_{B \neq A} [ \mc{J}_{BA}^A(\omega) P_A(t) + \mc{J}_{AB}^A(\omega) P_B(t) ].
\end{equation}
With this, $\mc{D}_A(\omega, t) \: d\omega$ becomes the rate of dissipation through the frequency window $[\omega, \omega + d\omega]$ at time $t$ for the vibrational modes coupled to molecule $A$. For future reference, we also define the cumulative dissipation density as
\begin{equation}\label{eq:diss_acc}
    \mc{E}_A(\omega, t) = \int_0^t \mc{D}_A(\omega, t') \: dt',
\end{equation}
which yields the total dissipated energy at $t$ when integrated over the frequency axis. We note that the subscript $A$ in $\mc{D}_A(\omega, t)$ and $\mc{E}_A(\omega, t)$ will be omitted when it is not needed for clarity.

The frequency-dependent profiles $\{\mc{J}_{BA}^C(\omega)\}$ in \eq{eq:diss_dens} are given by
\begin{equation}\label{eq:DSD}
    \mc{J}_{BA}^C(\omega) = \frac{2|V_{AB}|^2}{\hbar^2} \frac{J_{CC}(\omega)}{\omega} \mc{I}_{BA}(\omega)
\end{equation}
where $C$ is either $A$ or $B$. This quantity, which was called ``dissipative spectral density'' in our earlier work,\cite{Kim2021} captures how the energy dissipated by the transfer of state population from $\ket{A}$ to $\ket{B}$ is distributed among the bath modes coupled to $\ket{C}$. Moreover, as $J_{CC}(\omega)/\omega$ is the density of reorganization energy along the frequency domain [\eq{eq:spd_reorg_relation}], the quantity $\mc{I}_{BA}(\omega)$ in \eq{eq:DSD} reflects the ability of the bath to induce dissipation per unit reorganization energy. For this reason, $\mc{I}_{BA}(\omega)$ was named as ``dissipative potential'' and is expressed as
\begin{equation}\label{eq:Idiss}
    \begin{split}
    \mc{I}_{BA} (\omega) &= \ti{Re} \int_0^\infty \exp \bigg( - \frac{it' (E_B - E_A + \Lambda_{AA} + \Lambda_{BB})}{\hbar} - g_{AA}(t') - g_{BB}(t') \bigg) \\
    &\times \bigg[ \cos(\omega t') - i\coth \bigg( \frac{\beta \hbar \omega}{2} \bigg) \sin(\omega t') \bigg] \: dt'.
    \end{split}
\end{equation}
An interesting consequence of Eqs.~(\ref{eq:diss_dens})--(\ref{eq:Idiss}) is that, for two-level subsystems, the dissipation by both molecules become completely identical when $J_A(\omega) = J_B(\omega)$. This is because $\mc{J}_{BA}^A(\omega) = \mc{J}_{BA}^B(\omega)$ and $\mc{J}_{AB}^A(\omega) = \mc{J}_{AB}^B(\omega)$ under such a condition [\eq{eq:DSD}], and inserting these relations in \eq{eq:diss_dens} gives $\mc{D}_A(\omega, t) = \mc{D}_B(\omega, t)$. The validity of this corollary will be scrutinized in \stn{subsection:asymm}.

\subsubsection{Spin-Boson Model}\label{subsubsection:sbm}
In this model, a two-level subsystem is coupled to a single group of bath modes in an anti-correlated fashion. If we denote the two subsystem states as $\ket{+}$ and $\ket{-}$, the Hamiltonian components $\hat{H}_\ti{sub}$ and $\hat{H}_\ti{int}$ are written as
\begin{equation}\label{eq:H_sub_sb}
    \hat{H}_\ti{sub} = \frac{E}{2} \hat{\sigma}_z + V \hat{\sigma}_x,
\end{equation}
\begin{equation}\label{eq:H_int_sb}
    \hat{H}_\ti{int} = \hat{\sigma}_z \otimes \bigg[ \sum_j \bigg( - \omega_j^2 d_j \hat{x}_j + \frac{\omega_j^2 d_j^2}{2} \bigg) \bigg],
\end{equation}
where we have used the Pauli spin operators $\hat{\sigma}_x = \ket{+}\bra{-} + \ket{-}\bra{+}$ and $\hat{\sigma}_z = \ket{+}\bra{+} - \ket{-}\bra{-}$.


\section{Markovian Dissipation}\label{section:accuracy}
We now assess the accuracy of MQME-D by comparing its predictions with those obtained by using the numerically exact HEOM method.\cite{Tanimura1990,Tanimura2020} Our aim is to conduct a quantitative and systematic study on the extent to which the second-order perturbation theory and Markov approximation underlying MQME-D affect its reliability. In analogy with MQME-D, we call the decomposition of the dissipation into individual bath components based on HEOM as ``HEOM-D.'' In HEOM-D, the dissipation induced by a particular bath mode is indirectly elucidated by introduction of an additional ``probe mode,'' whose Hamiltonian closely resembles that of the mode we are interested in.\cite{Kim2022} The introduction of the probe mode enables us to extract the dissipation by using the conventional protocol based on the extended subsystem,\cite{O'Reilly2014,Novoderezhkin2017,Bennett2018} without disturbing the analytical BSD required to construct HEOM. Details of this exact method are included in \app{section:HEOM_D}.

For definitiveness, we focus on the dynamics of the local bath model (\stn{subsubsection:lbm}) whose subsystem just consists of two states which we will denote as $\ket{1}$ and $\ket{2}$. This model can be used to describe the dynamics within the single-excitation manifold of a molecular dimer, by mapping $\ket{1}$ ($\ket{2}$) onto the instance in which molecule 1 (2) is in its excited state while molecule 2 (1) remains in the ground state. When the electronic coupling arises from weak dipole-dipole interaction between the chromophores, the migration of excitation between them is referred to as Förster resonance energy transfer (FRET).\cite{Forster1959}

\subsection{Simulation Procedure for Molecular Dimer Model}\label{subsection:sim_prod}
We use Planck atomic unit system for which $\hbar = k_\ti{B} = 1$. Table~\ref{tab:table1} summarizes the simulation conditions employed for the molecular dimer model. We examine the accuracy of MQME-D by either varying $\Lambda$ while keeping the temperature constant as $T=1.0$ (conditions (i)--(iv) in Table~\ref{tab:table1}) or varying $T$ for a fixed reorganization energy $\Lambda = 0.2$ (conditions (ii), (v), and (vi) in Table~\ref{tab:table1}). From now on, we will refer these two sets of simulation conditions as ``const-$T$'' and ``const-$\Lambda$'' series, respectively. For each simulation condition in Table~\ref{tab:table1}, we also studied the effect of the energy gap $\Delta E = E_1 - E_2$ on the accuracy by varying $\Delta E$ as 0, 1 and 2, generating a total of 18 different simulation conditions. We fix $V_{12} = 0.25$ for the electronic coupling and $\omega_\ti{c} = 0.5$ for the cutoff frequency of the Drude-Lorentz BSDs [\eq{eq:DL_BSD}] that couple to the chromophores. We also assume that the electronic excitation is initially localized at molecule 1 unless noted otherwise.

For MQME and MQME-D simulations, each BSD was discretized into 2000 harmonic oscillator modes by using the scheme described in \app{section:BSD_disc}, which was originally reported in Ref.\citenum{Wang1999}. We used $\omega_\ti{max} = 15$ as the upper limit of frequency, which recovered 97.9\% of the pristine reorganization energy of the analytical BSD. The time integrals required to calculate electronic [\eq{eq:FRET_elec}] and dissipation [\eq{eq:DSD}] rate constants were evaluated by using the trapezoidal method with an integration grid size of 0.01 and the upper limit of integration of $5 \times 10^3$. The rate equations for electronic populations [\eq{eq:FRET_pop}] and dissipation [\eq{eq:diss_dens}] were propagated by using the fourth-order Runge-Kutta method with the time step of 0.01.

Numerically accurate benchmarks for dissipation were extracted by combining HEOM with the approach described in Ref.~\citenum{Kim2022}, along with the efficient low-temperature correction scheme recently reported in Ref.~\citenum{Fay2022}. Table~\ref{tab:table1} lists the number of hierarchy tiers $N_\ti{hier}$, the number of Matsubara low-temperature correction terms $N_\ti{Matsu}$, and Huang-Rhys (H-R) factor of the probe mode $s_\ti{bp}$ [\eq{eq:HR_probe}] used for the individual simulation conditions. For HEOM calculations, larger $N_\ti{hier}$ and $N_\ti{Matsu}$ are required for stronger subsystem-bath interaction and lower temperature, respectively, while $s_\ti{bp}$ is exclusively used for HEOM-D and must be small enough to satisfy the weak-coupling limit [\eq{eq:diss_bathmode_HR}]. We scanned the frequency of the probe mode in the range of $[0.1, 3.0]$ with a constant spacing of 0.05. For all data points $\omega \ge 0.2$, the number of vibrational quantum states used for describing the probe mode\cite{Kim2022} was determined to make the initial bath density captures 99.9\% of the total Boltzmann population. In the case of $\omega < 0.2$, the above criterion was slightly relaxed to 99.0\% to cope with the steeply increasing computational burden of HEOM-D as $\omega$ decreases (\stn{section:efficiency}).

The steady state limits ($t \rightarrow \infty$) were practically chosen as some finite time $t_\ti{sim}$, which is differently defined for each simulation condition by visually inspecting the evolution of electronic population. The reduced density matrix (RDM) of the subsystem and the auxiliary density matrices (ADMs) in HEOM are propagated by using the adaptive RKF45 integrator.\cite{Fehlberg1969} The error function used for tuning the time step was determined as how much the trace of the RDM deviates from unity, which must be maintained as nought in the exact dynamics. To further improve the stability of the calculation near the steady state, we also prevented the time step from increasing beyond a pre-determined maximum value $\Delta t_\ti{max}$, which is also listed in Table~\ref{tab:table1} for all simulation conditions.

\begin{table*}[htbp]
\caption{ \label{tab:table1} 
Summary of the simulation conditions used for the molecular dimer model. The quantities $\Lambda$ and $T$ define the subsystem-bath interaction. The other parameters specify the HEOM procedure. Each of the 6 simulation conditions in the table was combined with three different values of the energy gap $\Delta E = E_1 - E_2$ to yield 18 different conditions in total.
}

\begin{ruledtabular}
\begin{tabular}{lcccccc}
Simulation condition & (i) & (ii) & (iii) & (iv) & (v) & (vi) \\ \hline
Reorganization energy ($\Lambda$) & 0.05 & 0.2 & 1.0 & 2.0 & 0.2 & 0.2 \\
Temperature ($T$) & 1.0 & 1.0 & 1.0 & 1.0 & 0.5 & 0.25 \\
Maximum time step ($\Delta t_\ti{max}$) & 0.02 & 0.1 & 0.05 & 0.05 & 0.1 & 0.1 \\
Number of hierarchy tiers ($N_\ti{hier}$) & 4 & 7 & 10 & 13 & 7 & 7 \\
Number of Matsubara terms ($N_\ti{Matsu}$) & 30 & 30 & 30 & 30 & 100 & 100 \\
H-R factor of the probe mode ($s_\ti{bp}$) & $2 \times 10^{-6}$ & $1 \times 10^{-5}$ & $1 \times 10^{-5}$ & $1 \times 10^{-5}$ & $1 \times 10^{-5}$ & $1 \times 10^{-5}$
\end{tabular}
\end{ruledtabular}
\end{table*}

When the dynamics approaches the steady state, we observe that the total amount of dissipated energy computed with HEOM-D artificially increases at a constant rate for HEOM-D simulations even after the electronic dynamics has become stationary, violating energy conservation. Such a spurious behavior tends to become more severe when $\Lambda$ increases. A detailed numerical analysis of the drift is presented in the Supplementary Material for the case of $\Lambda = 1.0$ [condition (iii)]. As shown, the drift cannot be eliminated by simply increasing $N_\ti{hier}$ or $N_\ti{Matsu}$, or decreasing $\Delta t_\ti{max}$ or $s_\ti{bp}$. We therefore conclude that the dissipation does not arise from insufficient numerical convergence. In our results, we removed this artificial drift by applying a procedure similar to the one described in Ref.~\citenum{Kim2021} [Eq.~(S1) in the Supplementary Information]. This correction scheme assumes the linearity of the drift throughout the entire simulation, whose validity is also discussed in the Supplementary Material. Figure~S4 of the Supplementary Material demonstrates that the corrected dissipation is robust to the choice of the simulation parameters, as long as the state population achieves convergence.

\subsection{Electronic Dynamics}\label{subsection:electronic}

\begin{figure}[htbp]
\includegraphics{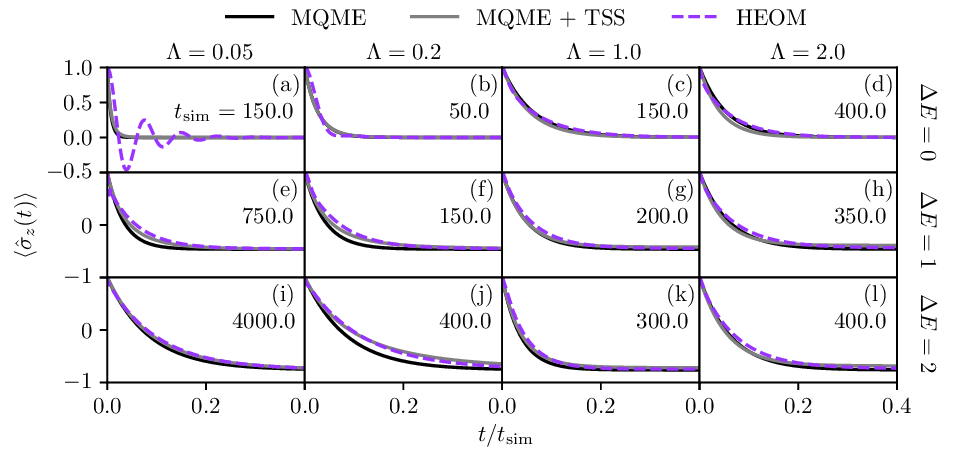}
\caption{\label{fig:fig1} Comparison between the dynamics of the state population represented as the population inversion $\langle \hat{\sigma}_z \rangle$, calculated by MQME (solid black) and HEOM (dashed purple) for the const-$T$ series ($T = 1.0$). The results calculated by combining MQME with the time scale separation (gray) is also presented, which will be discussed in \stn{subsubsection:dimer}.}
\end{figure}

\begin{figure}[htbp]
\includegraphics{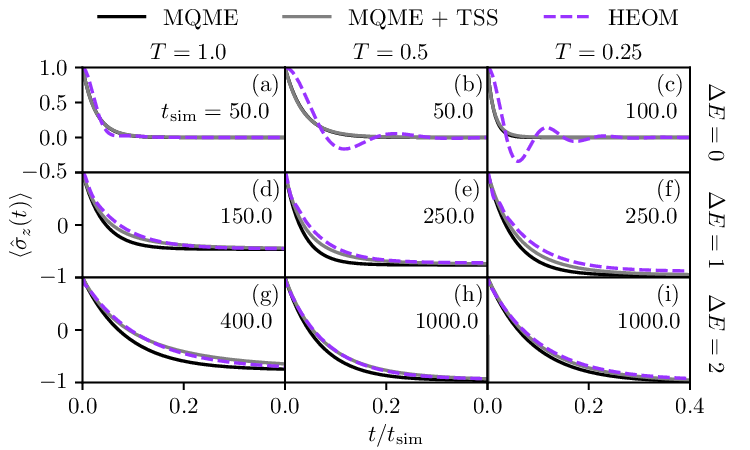}
\caption{\label{fig:fig2} Same as in \fig{fig:fig1}, but for the const-$\Lambda$ series ($\Lambda = 0.2$).}
\end{figure}

We first assess the reliability of MQME on describing the evolution of the state populations, which is the prerequisite for accurately calculating the dissipation. Before we present the results from the simulation, it is worthwhile to address how the applicability of MQME depends on the parameters related to the dynamics. According to Sec.~II A of Paper I,\cite{Kim2024} MQME is based on the assumption that the inter-state coupling is sufficiently weak so that it can be treated to second-order in perturbation. This implies that MQME will become accurate either when $|\Delta E| \gg |V_{12}|$ in $\hat{H}_\ti{sub}$, or when $\hat{H}_\ti{int}$ induces strong thermal fluctuation arising from high $T$ or large $\Lambda$.\cite{MontoyaCastillo2015}

We now analyze the accuracy of MQME based on the results obtained from the simulations. Figures \ref{fig:fig1} and \ref{fig:fig2} visualize the time-dependent population inversion $\langle \hat{\sigma}_z (t) \rangle = P_1(t) - P_2(t)$ calculated for the const-$T$ and const-$\Lambda$ series, respectively, by using both MQME and HEOM. For \fig{fig:fig1}, most of the cases [\fig{fig:fig1}(c)--(l)] show good agreement between MQME and HEOM results, even though MQME slightly overestimates the rate of population transfer when $\Delta E \ge 1$ [\fig{fig:fig1}(e)--(l)]. The most noticeable disagreement between MQME and HEOM is observed for $\Lambda = 0.05$ and $\Delta E = 0$ [\fig{fig:fig1}(a)], for which HEOM shows oscillations in $\langle \hat{\sigma}_z (t) \rangle$ while MQME does not. Such oscillations occur when the eigenstates of $\hat{H}_\ti{sub}$ undergo significant delocalization due to low $|\Delta E|$, and the coherence between these eigenstates can persist for relatively long time due to weak subsystem-bath interaction [\fig{fig:fig1}(a) and (b)]. It is easily expected that the accuracy of MQME will deteriorate under such conditions if we recall the discussions made above. The observed lack of oscillation actually shows a fundamental limitation of the MQME, as it can only predict monotonous, exponential dynamics. This can be explicitly shown by solving the coupled differential equations for the state populations [\eq{eq:FRET_pop}] for our dimer model. As a result, we can derive the analytical expression for the population inversion as
\begin{equation}
    \langle \hat{\sigma}_z (t) \rangle = \langle \hat{\sigma}_z (\infty) \rangle + \big[ \langle \hat{\sigma}_z (0) \rangle - \langle \hat{\sigma}_z (\infty) \rangle \big] e^{-(K_{12} + K_{21}) t}.
\end{equation}
As the rate constants $K_{AB}$ calculated by \eq{eq:FRET_elec} are always real, the state populations in MQME can only undergo a simple exponential decay without any oscillations.

The results for const-$\Lambda$ series in \fig{fig:fig2} exhibit a similar trend as observed in \fig{fig:fig1}. Namely, MQME accurately describes the population transfer when $\Delta E \ge 1$ [\fig{fig:fig2}(d)--(k)] with slight overestimations in the rate, although the performance becomes poor for the cases with $\Delta E = 0$ [\fig{fig:fig2}(a)--(c)]. The discrepancy between MQME and HEOM becomes more provoked as the temperature decreases, due to the reduced strength of the thermal fluctuation induced by the subsystem-bath interaction.

\subsection{Dissipation Dynamics}\label{subsection:dissipation}

\begin{figure}[htbp]
\includegraphics{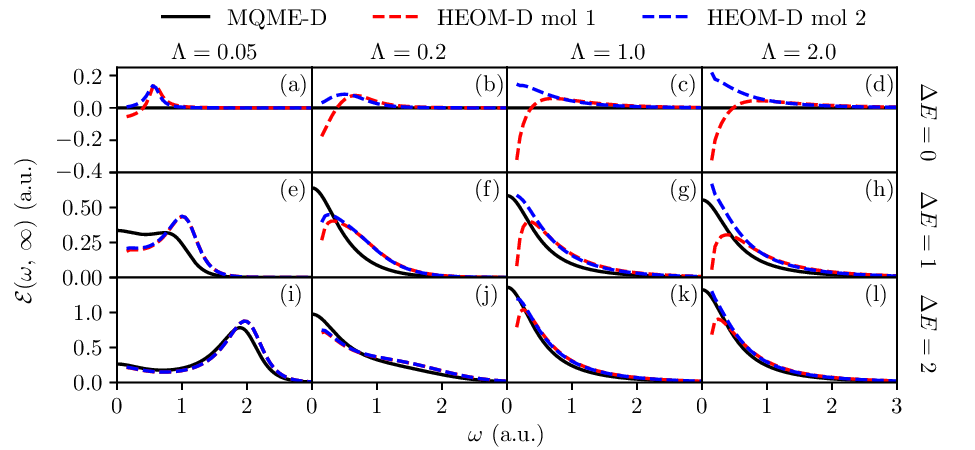}
\caption{\label{fig:fig3} Steady-state cumulative dissipation density $\mc{E}(\omega, \infty)$ calculated for const-$T$ series by using both MQME-D and HEOM-D. For MQME-D, the results for both molecules are identical and therefore plotted as a single profile. For HEOM-D, the results for the two molecules are plotted separately.}
\end{figure}

Having confirmed the reliability of MQME for the electronic dynamics, we now examine the dissipation. \Fig{fig:fig3} shows how the dissipated energy is distributed along the frequency axis for const-$T$ series, calculated by using both MQME-D and HEOM-D. As we have seen in \stn{subsection:equations}, MQME-D predicts the dissipation into both molecules to be exactly identical when $J_1(\omega) = J_2(\omega)$.\cite{Kim2021} With HEOM-D we can now put this statement under a close inspection, which was not attempted in Ref.~\citenum{Kim2021} due to the limited accuracy of the benchmark simulation method employed therein. We therefore avoid the redundancy by plotting the dissipation for only one molecule for MQME-D simulations, while separately plotting for either molecules for HEOM-D simulations.

The dissipation can be conveniently visualized in the frequency domain by using the cumulative dissipation density defined in \eq{eq:diss_acc}. \Fig{fig:fig3} shows $\mc{E}(\omega, \infty)$ for const-$T$ series obtained by both MQME-D and HEOM-D. When $\Delta E = 0$ [\fig{fig:fig3}(a)--(d)], MQME-D predicts vanishing dissipation for all $\omega$, although HEOM-D results clearly demonstrate net energy transfer from the subsystem to bath for all four values of $\Lambda$. Such an incorrect behavior of MQME-D arises because $E_A = E_B$ makes $\mc{J}_{BA}^A(\omega) = \mc{J}_{AB}^A(\omega)$ [\eqs{eq:DSD}{eq:Idiss}], while the detailed balance condition also imposes $\mc{J}_{BA}^A(\omega) = - \mc{J}_{AB}^A(\omega)$ (Sec.~II~C of Paper I\cite{Kim2024}). These two conditions can be simultaneously met only when both $\mc{J}_{AB}^A(\omega)$ and $\mc{J}_{BA}^A(\omega)$ vanishes for all $\omega$, leading to the absence of dissipation as shown in \fig{fig:fig3}(a)--(d). In reality, however, energy is dissipated from the subsystem to the bath due to the population relaxation between the eigenstates of $\hat{H}_\ti{sub}$. While such an aspect is naturally incorporated in HEOM-D, MQME-D cannot handle the effects arising from delocalized eigenstates due to its focus on the projected system density under second-order perturbation (Sec.~II~A of Paper I\cite{Kim2024}).

When $\Delta E$ becomes larger, we observe much better qualitative agreement between the predictions of MQME-D and HEOM-D. Of particular interest are the cases with $\Delta E = 2$, for which MQME-D exhibits semi-quantitative accuracy [\fig{fig:fig3}(i)--(l)] with slight overestimation (underestimation) of the dissipation when $\omega$ is small (large). With a relatively small reorganization energy of $\Lambda = 0.05$ [\fig{fig:fig3}(i)], we observe that a substantial portion of the dissipation occurs through the region around $\hbar \omega = \Delta E$, which can be related to the vibronic resonance. Increasing $\Lambda$ makes the contribution of this channel gradually disappear while the dissipation becomes more concentrated near $\omega = 0$. However, the MQME-D calculations predict that $\mc{E}(\omega, \infty)$ monotonously increases as $\omega$ approaches zero, which contradicts the steep drops observed in the HEOM-D counterparts [\fig{fig:fig3}(k) and (l)]. Our previous study\cite{Kim2021} clarified that the discrepancy at the low frequency is caused by the Markov approximation behind MQME-D, which neglects the quasi-static nature of the bath modes that delays their participation in the dynamics.

\begin{figure}[htbp]
\includegraphics{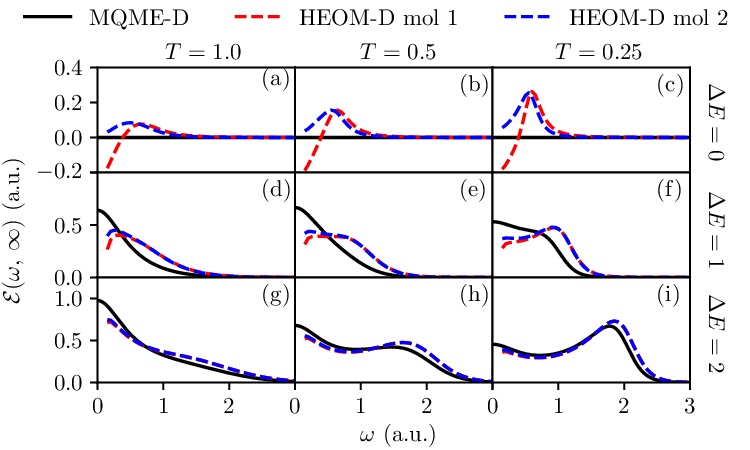}
\caption{\label{fig:fig4} Same as in \fig{fig:fig3}, but for the const-$\Lambda$ series.}
\end{figure}

Finally, we inspect the effect of temperature on dissipation by looking into \fig{fig:fig4}, which summarizes $\mc{E}(\omega, \infty)$ for const-$\Lambda$ series calculated by both MQME-D and HEOM-D. As in \fig{fig:fig3}, MQME-D erroneously predicts zero dissipation for $\Delta E = 0$ [\fig{fig:fig4}(a)--(c)]. Nevertheless, the accuracy of MQME-D increases with $\Delta E$ and reaches semi-quantitative level for $\Delta E = 2$ [\fig{fig:fig4}(g)--(i)]. Apparently, lowering $T$ from 1.0 to 0.25 enhances the influence of the vibronic resonance in dissipation, which is due to the reduction of the thermal fluctuation induced by the subsystem-bath interaction. Similar to what we have seen from \fig{fig:fig3}, the dissipation calculated by MQME-D tends to be concentrated toward slightly lower frequency compared to the HEOM-D results. Again, this is because MQME-D is based on Markov approximation and therefore overestimates the contribution of the low-frequency bath modes on the dissipation.


\subsection{Asymmetry of the Dissipation between Molecules}\label{subsection:asymm}
As we have stated earlier, MQME-D predicts the dissipation into the two molecules in a dimer to be exactly identical.\cite{Kim2021} According to Figs.~\ref{fig:fig3} and \ref{fig:fig4}, this prediction is satisfied for large $\omega$ and small $\Lambda$. However, HEOM-D calculations also show that some asymmetry do exist between $\mc{E}_1(\omega, \infty)$ and $\mc{E}_2(\omega, \infty)$, especially when $\Delta E$ is small [Figs.~\ref{fig:fig3}(a)--(d)] or $\Lambda$ is relatively large [Figs.~\ref{fig:fig3}(k) and (l)]. The most dramatic cases are the conditions with $\Delta E = 0$ [Figs. 3(b)-3(d) and Figs. 4(a)-4(c)], where the directions of dissipation in the two molecules become completely opposite for low-frequency bath modes. This shows that the low-frequency modes of molecule 1 only lose energy from the reorganization, while those of molecule 2 actively absorb energy from the subsystem. It would be meaningful to conduct a further inspection about the origin of this asymmetry to clarify how it is connected to the approximations underlying MQME-D.

As the asymmetry becomes more pronounced when $\omega$ is small, we can speculate that the phenomenon is related to non-Markovianity. A possible source of the non-Markovianity in our dimer system is the non-equilibrium motion of the bath triggered by Franck-Condon transition to the excited state PES. To elaborate further on this phenomenon, we separately examine the behavior of the bath density in MQME and HEOM dynamics. In Sec.~II~B of Paper I,\cite{Kim2024} we explained that MQME focuses only on the projected component of the density matrix for the system $\hat{\rho}(t)$
\begin{equation}
    \hmc{P}\hat{\rho}(t) = \sum_A \bigg( P_A(t) \ket{A} \bra{A} \otimes \frac{\exp(-\beta \hat{h}_A)}{\tr_\ti{b} [ \exp(-\beta \hat{h}_A) ]} \bigg),
\end{equation}
where $\hat{h}_A = \bra{A} \hat{H} \ket{A}$ and $\tr_\ti{b}$ denotes the trace over the subspace spanned by the bath DOFs. If we combine this condition with the explicit expressions for $\hat{H}$ and its components [Eqs.~(\ref{eq:H})--(\ref{eq:H_int})], we can observe that the molecular vibrational modes in MQME dynamics are always in the thermal equilibrium associated with the PES of the relevant electronic state. As the shape of the PES for the ground and excited states are identical in our Hamiltonian model [\eqs{eq:H_bath_loc}{eq:H_int_loc}], the excited state bath density is identical to the ground state density except its center is shifted to the minimum of the excited state PES. Under such a condition, the relative dissipation rates for the vibrational modes are solely determined by the distance between the minima of ground and excited state PESs. Because these distances are identical for both molecules when their spectral densities [\eq{eq:BSD}] are the same, the dissipation by the two molecules must be symmetric as already proven in \stn{subsection:equations}.

In contrast to MQME, HEOM assumes that the initial system density to be in the direct product form $\hat{\rho}(0) = \hat{\sigma}(0) \otimes \hat{R}_\ti{g}$, where $\hat{\sigma}(0)$ is the initial subsystem density and 
\begin{equation}
    \hat{R}_\ti{g} = \frac{\exp(-\beta \hat{H}_\ti{bath})}{\tr_\ti{b} [ \exp(-\beta \hat{H}_\ti{bath}) ]}
\end{equation}
is the equilibrium bath density on the PES of the electronic ground state.\cite{Tanimura2014} Because we have set the initial electronic populations as $P_1(0) = 1$ and $P_2(0) = 0$, the bath modes coupled to molecule 1 undergo Franck-Condon excitation and start oscillating in the excited state PES while those coupled to molecule 2 do not. We can easily expect that such a difference in the dynamics of the bath density would cause asymmetry in the dissipation by affecting microscopic energy flows between vibronic quantum states.

\begin{figure}[htbp]
\includegraphics{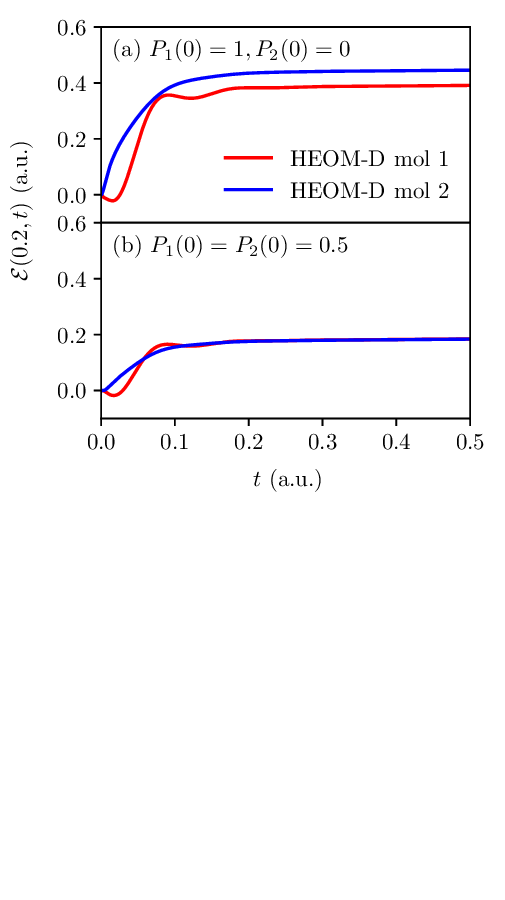}
\caption{\label{fig:fig5} Densities of accumulated 
dissipation at $\omega = 0.2$, separately obtained for both molecules by using HEOM-D under simulation condition (iii) (Table \ref{tab:table1}) for $\Delta E$ = 2. Two different initial electronic populations were employed to demonstrate that the asymmetry between the two molecules are related to non-equilibrium motion of the bath modes.}
\end{figure}

To corroborate our explanation further, we directly calculate and visualize how the asymmetry in the dissipation is affected by the initial electronic populations. For this purpose, we have specifically chosen the case of $\omega = 0.2$ and $\Lambda = 1.0$ which exhibits a noticeable difference between $\mc{E}_1(\omega, \infty)$ and $\mc{E}_2(\omega, \infty)$ [\fig{fig:fig3}(k)]. We calculated the time-dependent dissipation $\mc{E}(0.2, t)$ for two different initial conditions: (i) $P_1(0) = 1$, $P_2(0) = 0$ and (ii) $P_1(0) = P_2(0) = 0.5$. The results plotted in \fig{fig:fig4} illustrate that the asymmetry under the original initial condition [\fig{fig:fig5}(a)] disappears as expected when we induce the same amount of Franck-Condon excitation for both molecules by setting their initial electronic populations as equal [\fig{fig:fig5}(b)]. Hence, it supports our claim that the asymmetry in the dissipation is indeed linked to the difference between the non-equilibrium motion of the bath. Intriguingly, although $\mc{E}_1(0.2, t)$ in the early stage of the dynamics clearly exhibits oscillations arising from nuclear motions, such a feature is not visible in $\mc{E}_2(0.2, t)$ even with some amount of excitation initially residing in molecule 2 [\fig{fig:fig5}(b)].

\section{Non-Markovian Dissipation via Time Scale Separation (TSS)}\label{section:nonMarkovian}
In this section, we conduct extensive tests on MQME-D+TSS. \Stn{subsection:intro_tss} introduces the principles of TSS. \Stn{subsection:diss_tss} uses MQME-D+TSS to compute the dissipation in different types of Hamiltonian models with a wide range of simulation parameters. The results are then compared with HEOM-D to appraise the accuracy of our theory.

\subsection{Introduction to Time Scale Separation}\label{subsection:intro_tss}
In \stn{subsection:dissipation}, we have seen that one source of inaccuracy in MQME-D is the Markov approximation. This suggests that the reliability of MQME-D may be increased if we can somehow include non-Markovianity in the simulation. This task can be accomplished by employing TSS,\cite{Berkelbach2012,MontoyaCastillo2015} which divides the BSD $J(\omega)$ into ``slow'' $J_\ti{slow}(\omega)$ and ``fast'' $J_\ti{fast}(\omega)$ components, respectively. The desired non-Markovianity is introduced in the simulation by prohibiting the bath modes in $J_\ti{slow}(\omega)$ from directly participating in the dynamics. This is practically achieved by defining the BSD components as
\begin{subequations}\label{eq:BSD_TSS}
\begin{equation}
    J_\ti{slow}(\omega) = S(\omega, \omega^*) J(\omega),
\end{equation}
\begin{equation}
    J_\ti{fast}(\omega) = [1 - S(\omega, \omega^*)] J(\omega).
\end{equation}
\end{subequations}
Here, $S(\omega, \omega^*)$ is the splitting function which monotonically decays as $\omega$ increases, with the speed of decay controlled by the cutoff frequency $\omega^*$. In our simulations, by following the similar approach as in Ref.~\citenum{MontoyaCastillo2015}, we use the splitting function of the form
\begin{equation}\label{eq:f_sp}
    S(\omega, \omega^*) = \left\{
        \begin{array}{cc}
            \eta [1 - (\omega/\omega^*)^2]^2, & \omega < \omega^*, \\
            0, & \omega \ge \omega^*,
        \end{array}
        \right.
\end{equation}
where we have added an extra scaling factor $\eta < 1$ to the original expression to ensure the numerical convergence of the improper integrals [\eqs{eq:FRET_elec}{eq:Idiss}].

We now treat $J_\ti{slow}(\omega)$ as a source of static disorder, and modulate the state energies of the subsystem $\{E_A\}$ by adding Gaussian random noise whose standard deviation is
\begin{equation}\label{eq:stdev_TSS}
    \sigma_\ti{slow} = \frac{1}{\hbar} \int_0^\infty J_\ti{slow}(\omega) \coth  \bigg( \frac{\beta \hbar \omega}{2} \bigg) \: d\omega,
\end{equation}
while $J_\ti{fast}(\omega)$ governs the dissipation in individual realizations of the disorder. The final result is calculated by averaging over sufficiently large number of realizations. Each noise trajectory follows MQME whose BSD is $J_\ti{fast}(\omega)$, and therefore exhibits the characteristics of the bare MQME such as exponential time-dependence and detailed balance $P_A(\omega) / P_B(\omega) = \exp[\beta(E_B - E_A)]$. However, because the static disorder induces variation in the state energies, such properties are not satisfied after averaging over the entire set of trajectories.

\subsection{Dissipation Dynamics}\label{subsection:diss_tss}
We now benchmark the accuracy of MQME-D+TSS against different types of model Hamiltonians for open quantum systems. In \stn{subsubsection:dimer}, we first explore the performance of MQME-D + TSS for the molecular dimer model which was already employed to benchmark the original MQME-D in \stn{section:accuracy}. \Stn{subsubsection:sb_bo} applies MQME-D to the spin-boson Hamiltonian with Brownian oscillator BSD.

\subsubsection{Molecular Dimer}\label{subsubsection:dimer}

\begin{figure}[htbp]
\includegraphics{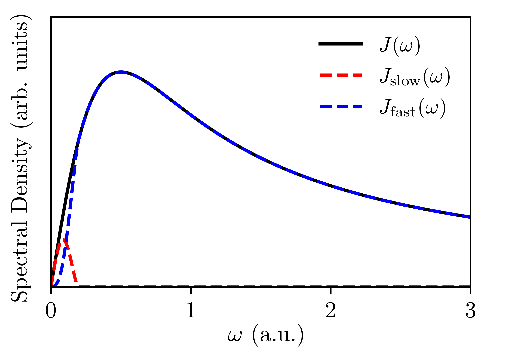}
\caption{\label{fig:fig6} Split of the Drude-Lorentz BSD [\eq{eq:DL_BSD}] into slow and fast components by employing \eqs{eq:BSD_TSS}{eq:f_sp} with $\eta = 0.99$ and $\omega^* = 0.2$.}
\end{figure}

We examine the same molecular dimer model used in \stn{subsection:dissipation}, defined by the local bath Hamiltonian [Eqs.~(\ref{eq:H_sub}), (\ref{eq:H_bath_loc}), and (\ref{eq:H_int_loc})]. We apply TSS by constructing the splitting function [\eq{eq:f_sp}] with $\eta = 0.99$ and $\omega^* = 0.2$, and averaging over $10^3$ noise trajectories for conditions (ii)--(vi) in Table~\ref{tab:table1}. For the condition (i) with $\Lambda = 0.05$, the number of trajectories was increased to $10^4$ to ensure numerical convergence. \Fig{fig:fig6} illustrates how TSS splits the Drude-Lorentz BSD used in our simulations. All other simulation procedures remain the same as those explained in \stn{subsection:sim_prod}.

We first examine the effect of TSS in population dynamics. \Fig{fig:fig1} plots $\langle \hat{\sigma}_z (t) \rangle$ obtained from both bare MQME and MQME+TSS obtained for const-$T$ series. It is observed that TSS modifies the rate of population relaxation, although the direction and magnitude of the influence show complicated dependence on both $\Lambda$ and $\Delta E$. Nevertheless, there are cases for which TSS substantially increases the accuracy of MQME. In particular, nearly perfect matches between MQME-TSS and HEOM are observed for Figs.~\ref{fig:fig1}(i) and (j). In Figs.~\ref{fig:fig1}(e) and (f), TSS also leads to better agreements with HEOM-D results by prolonging the relaxation. Nevertheless, the accuracy of MQME-D+TSS may become further improved by using different values of $\omega^*$. On the other hand, Figs.~\ref{fig:fig1} (c) and (d) also demonstrate that TSS does not always positively effect the accuracy. Similar trends are observed for const-$\Lambda$ series (\fig{fig:fig2}).

\begin{figure}[htbp]
\includegraphics{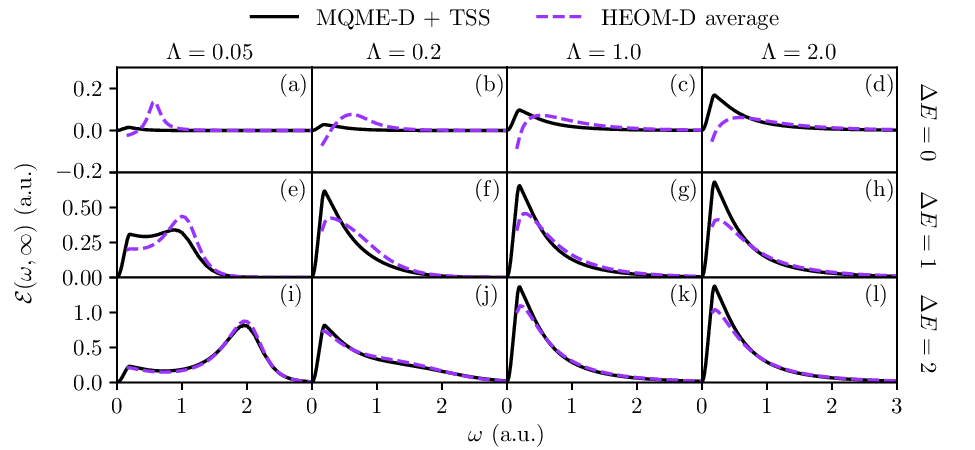}
\caption{\label{fig:fig7} Cumulative dissipation density at the steady state $\mc{E}(\omega, \infty)$, calculated for const-$T$ series by combining MQME-D+TSS with the cutoff frequency $\omega^* = 0.2$. The averages of the HEOM-D results (dashed purple) for both BSDs in \fig{fig:fig3} are plotted together to benchmark the accuracy of the results.}
\end{figure}

\begin{figure}[htbp]
\includegraphics{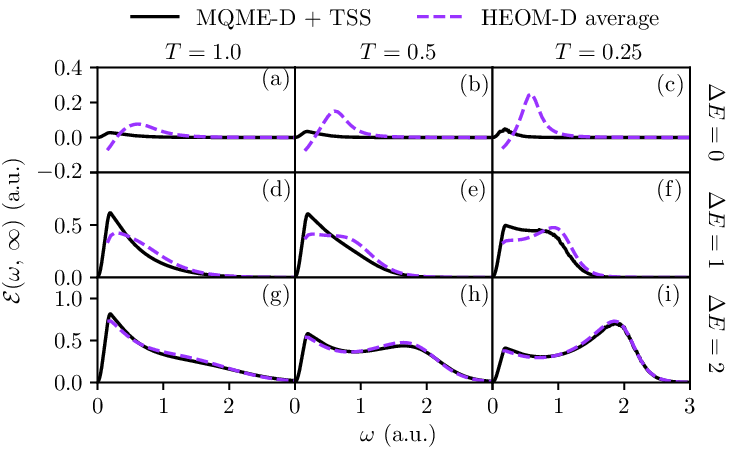}
\caption{\label{fig:fig8} Similar to \fig{fig:fig7}, but for const-$\Lambda$ series. In this case, the benchmark is the averages of the HEOM-D results in \fig{fig:fig4}.}
\end{figure}

We now examine the dissipation predicted by MQME-D+TSS to see how the added non-Markovianity affects the accuracy of the method. We note that TSS does not save the lack of asymmetry in MQME, as we are applying the same $S(\omega, \omega^*)$ to the BSDs of the two molecules. We therefore compare the resolved dissipation against the HEOM-D results averaged over the two molecules.

Figures $\ref{fig:fig7}$ and $\ref{fig:fig8}$ display $\mc{E}(\omega, t)$ calculated by MQME-D+TSS, along with the averaged $\mc{E}(\omega, t)$ obtained from HEOM-D simulations. Compared to the results obtained without TSS (Figs.~\ref{fig:fig2} and \ref{fig:fig3}), it is apparent that TSS improves the accuracy of MQME-D for both const-$T$ (\fig{fig:fig7}) and const-$\Lambda$ (\fig{fig:fig8}) series, especially for the cases with $\Delta E = 2$ [Figs.~\ref{fig:fig7}(i)--(l) and \ref{fig:fig8}(g)--(i)]. In particular, the agreement near $\omega = 0$ became remarkably better as $J_\ti{fast}(\omega)$ does not exhibit strong subsystem-bath interaction anymore in that region. The dissipation originally in this region is redirected toward higher frequencies, which alleviates the underestimation of $\mc{E}(\omega, t)$ near $\omega = 2$ by MQME-D. Such an effect is also pronounced for $\Delta E = 1$ [Figs.~\ref{fig:fig7}(e)--(h) and \ref{fig:fig8}(d)--(f)], although some discrepancy still remains.

Finally, for the homodimer case of $\Delta E = 0$ [Figs.~\ref{fig:fig1}(a)--(d) and \ref{fig:fig2}(a)--(c)], we observe some amount of dissipation in contrast to the vanishing dissipation in Figs.~\ref{fig:fig1} and \ref{fig:fig2}. This is because the individual noise trajectories exhibit nonzero $\Delta E$ due to the static disorder arising from $J_\ti{slow}(\omega)$. However, because MQME loses its reliability when $\Delta E = 0$, MQME-D is also not accurate enough to make meaningful predictions of dissipation.

Overall, we can expect MQME-D+TSS will accurately decompose the dissipated energy when MQME qualitatively accounts for the dynamics of state populations (large $\Delta E$ or $\Lambda$), and TSS leads to an additional increase in the accuracy at the quantitative level.

\subsubsection{Spin-boson Model with Brownian Oscillator Bath}\label{subsubsection:sb_bo}

We now test how MQME-D+TSS performs for the Hamiltonian models involving Brownian oscillator BSD $J_\ti{BO}(\omega)$ [\eq{eq:BO_BSD}]. We examine the dissipation induced by an underdamped bath mode whose frequency is tuned to achieve resonance with the subsystem. By modulating the reorganization energy $\Lambda$ and damping parameter $\gamma$, we conduct a systematic investigation on how the strength of subsystem-bath interaction and memory time of the bath affect the accuracy of our method.

\begin{figure}[htbp]
\includegraphics{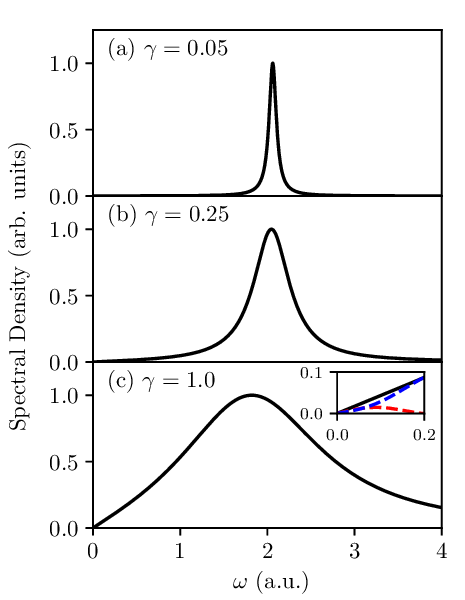}
\caption{\label{fig:fig9} Shapes of $J_\ti{BO}(\omega)$ used in the simulations of spin-boson model, for different values of the damping strength $\gamma$. The inset in panel (c) shows how TSS splits the BSD into $J_\ti{slow}(\omega)$ (dashed red) and $J_\ti{fast}(\omega)$ (dashed blue) when $\gamma = 1.0$.}
\end{figure}

\begin{table*}[htbp]
\caption{ \label{tab:table2} 
Summary of the simulation conditions used for the spin-boson model with Brownian oscillator bath. The first parameter is for both MQME and HEOM calculations, while the rest are specifically for HEOM. Each of the 3 simulation conditions in the table was combined with three different values of the damping constant $\omega_\ti{c}$ to yield 9 different conditions in total.}

\begin{ruledtabular}
\begin{tabular}{lccc}
Simulation condition & (i) & (ii) & (iii) \\ \hline
Reorganization energy ($\Lambda$) & 0.05 & 0.25 & 1.0 \\
Maximum time step ($\Delta t_\ti{max}$) & 0.01 & 0.05 & 0.05 \\
Number of hierarchy tiers ($N_\ti{hier}$) & 5 & 7 & 12 \\
Secondary $N_\ti{hier}$ for $\omega_\ti{c} = 0.05$ & 10 & 15 & 25 \\
H-R factor of the probe mode ($s_\ti{bp}$) & $2 \times 10^{-6}$ & $1 \times 10^{-5}$ & $1 \times 10^{-5}$
\end{tabular}
\end{ruledtabular}
\end{table*}

We simulate the dynamics of a spin-boson Hamiltonian defined by Eqs.~(\ref{eq:H_bath}), (\ref{eq:H_sub_sb}), and (\ref{eq:H_int_sb}). The subsystem parameters are $E = 2$ and $V = 0.25$. For the BSD, we use three different values of $\Lambda$ as 0.05, 0.25 and 1.0 as listed in Table~\ref{tab:table2}, and vary the damping strength $\gamma$ as 0.05, 0.25, and 1.0 (\fig{fig:fig9}) for each value of $\Lambda$ to make a total of 9 simulation conditions. The characteristic frequency of the BSD was set as $\omega_0 = 2.062$ to match the difference between the eigenenergies of $\hat{H}_\ti{sub}$. The temperature of the bath was kept constant as $T = 1$ for all simulation conditions.

For MQME and MQME-D simulations, the rate constants for population transfer and dissipation were calculated according to \eqs{eq:FRET_elec}{eq:diss_mode} by setting $d_{\pm j} = \pm d_j$ in \eq{eq:H_int}. The BSD was discretized into 5000 harmonic oscillator modes based on the scheme described in \app{section:BSD_disc} for all simulation conditions in Table~\ref{tab:table2} except the case of $\Lambda = 0.05$ and $\gamma = 0.05$, which required 20000 oscillators to guarantee numerical convergence. The upper limit of frequency $\omega_\ti{max}$ and the integration scheme were the same as what we used for the dimer model (\stn{subsection:sim_prod}). The TSS was applied by using the same cutoff $\omega^* = 0.2$ as in the Drude-Lorentz BSDs in the dimer model (\stn{subsubsection:dimer}), while the scaling factor $\eta$ was reduced from 0.99 to 0.6 due to the increased difficulty of achieving detailed balance condition for $J_\ti{BO}(\omega)$. The number of individual noise trajectories was always kept as 1000.

For HEOM and HEOM-D simulations, we implemented the Brownian oscillator BSD based on the efficient framework reported in Ref.~\citenum{Ikeda2020} and combined it with the perturbative low-temperature correction\cite{Fay2022} with $N_\ti{Matsu} = 30$. The depth of hierarchy $N_\ti{hier}$ was adjusted depending on the reorganization energy as listed in Table~\ref{tab:table2}. When $\gamma = 0.05$, deeper hierarchy was needed for the numerical convergence near $\omega = \omega_0$ due to the strong resonance arising from the subsystem-bath interaction. We scanned the frequency of the probe mode in the range of [0.2, 1.9) and (2.2, 3.0] with a constant spacing of 0.05, while a finer grid of 0.005 was used for the range of [1.9, 2.2] for capturing the detailed structure of $\mc{E}(\omega, t)$ near the resonance.

\begin{figure}[htbp]
\includegraphics{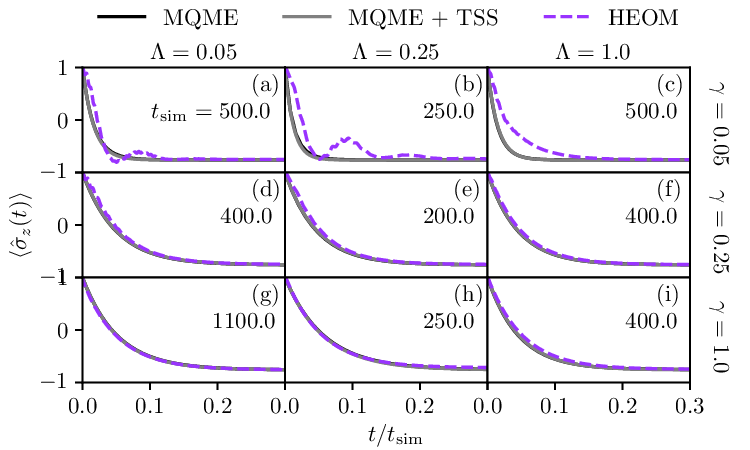}
\caption{\label{fig:fig10} Comparison between the evolution of population inversion $\langle \hat{\sigma}_z \rangle$ for the spin-boson model, calculated by MQME (solid black), MQME+TSS (solid gray), and HEOM (dashed purple).}
\end{figure}

\begin{figure}[htbp]
\includegraphics{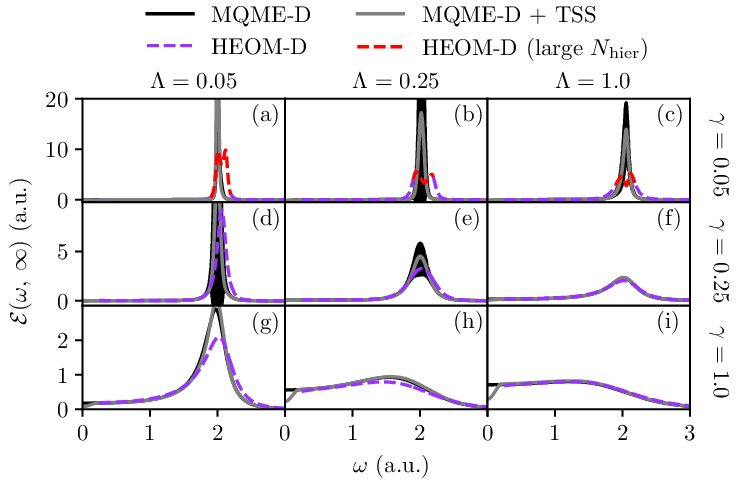}
\caption{\label{fig:fig11} Steady-state cumulative dissipation density $\mc{E}(\omega, \infty)$ calculated for the spin-boson model by using MQME-D (solid black), MQME-D+TSS (solid gray), and HEOM-D (dashed purple). For the conditions with $\gamma = 0.05$, we increased the hierarchy depth for HEOM-D (dashed red) in the frequency domain [1.9, 2.2] to guarantee the convergence under strong resonance between the subsystem and bath modes.}
\end{figure}

In \fig{fig:fig10}, we have presented the calculated $\langle \hat{\sigma}_z (t) \rangle$ for all 9 simulation conditions listed in Table~\ref{tab:table2}. The results for $\gamma = 0.05$ [Figs.~\ref{fig:fig10}(a)--(c)] show that it is challenging for MQME and MQME+TSS to describe highly non-Markovian character of the bath originating from the small damping. However, the agreement becomes much better as $\gamma$ increases, and show nearly quantitative match for [Figs.~\ref{fig:fig10}(d)--(i)]. Meanwhile, in contrast to the molecular dimer model coupled to Drude-Lorentz BSD (\stn{subsubsection:dimer}), there is almost no visible difference between MQME and MQME+TSS. This is because $J_\ti{BO}(\omega)$ takes relatively small value near $\omega = 0$, which makes $J_\ti{slow}(\omega)$ have only a minute contribution to the overall BSD [\fig{fig:fig9}(c)] in our simulation conditions.

We now examine \fig{fig:fig11} and discuss how the dissipation by the Brownian oscillator BSD looks like. For small ($\gamma = 0.05$) and intermediate ($\gamma = 0.25$) damping [Figs.~\ref{fig:fig11}(a)--(f)], HEOM-D results show that most of the dissipation occurs through the resonant channel around $\omega = 2$. What is interesting is that $\mc{E}(\omega, \infty)$ for $\gamma = 0.05$ [Figs.~\ref{fig:fig11}(a)--(c)] does not form a single peak as in the Brownian BSD (\fig{fig:fig9}) but instead a pair of peaks closely lying together. Such a structure arises from the interaction between the upper subsystem state $\ket{+}$ and the first excited state of the underdamped bath mode, as in the formation of a polaritonic state pair.\cite{Ribeiro2018,Ebbesen2023} Nevertheless, this behavior soon disappears as the effect of resonance is diluted due to the increased damping  [Figs.~\ref{fig:fig11}(d)--(i)]. For all panels in \fig{fig:fig11}, both MQME-D and MQME-D+TSS qualitatively reproduce the results from HEOM-D calculations. The predictability becomes better with increasing $\Lambda$ and $\gamma$ which enhances the adequacy of second-order perturbation and Markov approximation, respectively.

In contrast to the population dynamics for which TSS had virtually no effect, averaging over the TSS noise trajectories removes rapid oscillations in the dissipation, which appears near the resonance frequency ($\omega \approx 2$) in the bare MQME-D [Figs.~\ref{fig:fig11}(b)--(e)]. These oscillations arise from incomplete numerical convergence, and it is possible to mitigate them to some extent by extending the upper limit of integration in \eq{eq:diss_mode}. However, we found out that the convergence without TSS was extremely slow and the oscillations were still prevalent even after integrating up to $5 \times 10^5$ (100 times the limit used in the main calculations). Therefore, TSS offers a convenient way to achieve converged results under strong subsystem-bath resonance, which can be also straightforwardly parallelized by distributing the propagation and averaging procedure over multiple processors.

As seen from Figs.~\ref{fig:fig11}(d), (e), (g), and (h), MQME-D+TSS tends to overestimate the dissipation near the region where subsystem-bath resonance occurs. This is likely because the dissipated energy cannot return to the subsystem in MQME due to the second-order approximation, although such re-absorption of energy by subsystem does occur when the underdamped bath mode exerts strong subsystem-bath resonance.\cite{Kim2020}

\section{Computational Efficiency of MQME-D}\label{section:efficiency}
In this paper, we obtained exact decompositions of dissipation in our Hamiltonian models by combining the numerically exact HEOM method\cite{Tanimura1990,Ikeda2020} with a technique for extracting the statistics of a particular bath mode.\cite{Kim2022} However, such computations become exponentially more costly as the number of subsystem DOFs increases. Most of the other numerically exact simulation methods exhibit similar exponential scaling for propagating the subsystem RDM, although we are aware of a recently reported method that could potentially overcome this issue.\cite{Varvelo2021}

In addition, the computational cost of HEOM-D also depends on the temperature of the bath and the characteristic frequency of the harmonic bath mode whose dissipation we want to calculate. As explained in \app{section:HEOM_D}, HEOM-D extracts the dissipation by merging the subsystem with an extra bath mode which acts as a probe for monitoring the dynamics of the bath mode of interest. The probe mode must have the same frequency as the mode we want to monitor, and should be described by large enough number of quantum states to represent the thermal properties during the dynamics. Therefore, the number of required vibrational quantum states drastically increases as we reduce the energy spacing $\hbar \omega$ below $k_\ti{B} T$ and move toward zero. If $n$ quantum states are used to describe the probe mode, the cost of propagating the reduced density matrix and auxiliary density matrices of the extended subsystem would approximately depend on $O(n^3)$, as it involves matrix-matrix multiplications arising from commutators and anti-commutators. Moreover, the perturbative low-temperature correction described in Ref.~\citenum{Fay2022} requires diagonalizations of the super-operators represented in the Liouville space, whose cost would exhibit the dependence of $O(n^6)$. Such a steep growth of the computational burden was the reason why we always needed to terminate HEOM-D calculations at a certain lower limit for $\omega$.

On the other hand, the aforementioned issues pose much less difficulty for MQME-D. Namely, it is less problematic to apply MQME-D to large systems as the cost for evaluation of the dissipation rate constants and time propagation only grows as $O(N^2)$ where $N$ is the dimension of the subsystem, if we assume that all subsystem DOFs are coupled to an identical number of bath modes. In addition, MQME-D can be trivially parallelized by distributing the load of evaluating the rate constants and propagating TSS-based noise trajectories across multiple processors. This is in contrast to HEOM\cite{Johan2012,Noack2018} and tensor-train-based simulation methods\cite{Urbanek2016,Strathearn2018} whose equations-of-motion are usually densely coupled and therefore require substantial amounts of communication for propagation. Furthermore, the cost of evaluating the dissipation rate constants based on \eq{eq:diss_mode} does not depend on $\omega$, which allows us to conveniently access the dissipation in low-frequency region without any additional burden.

Finally, when one employs numerically exact simulation methods to calculate the dissipation into individual bath modes, only one mode is usually monitored at a time to keep the dimension of the subsystem RDM within a viable extent. As a result, the formation and propagation of the extended subsystem needs to be repeated for every bath mode to obtain a complete decomposition of the dissipation. On the other hand, provided that the rate constants have been already computed, MQME-D captures all information regarding the dissipation in a single propagation. Based on these observations, we expect MQME-D with TSS to have a promising utility in studying the role of individual bath modes to the quantum dynamics in large molecular systems, whose details cannot be easily accessed by numerically exact simulation methods.

\section{Conclusion}\label{section:Conclusion}
In this paper, we investigated the accuracy of MQME-D, a theoretical method that enables us to decompose the dissipation under MQME dynamics into contributions from individual bath modes. The theory was applied to multiple types of Hamiltonian models and the outcomes from the simulations were compared against numerically exact results provided by HEOM. We have demonstrated that the dissipation calculated by MQME-D offers a qualitatively correct view of the dissipated energy by individual bath modes. However, it can quantitatively differ from the exact results even in the limit where MQME exhibits accurate population dynamics. We have provided detailed arguments that support Markovian origin of the observed discrepancies, and demonstrated that the accuracy of the calculation is indeed significantly increased by inclusion of non-Markovianity via TSS. In the end, despite the inherent limitation arising from second-order perturbation approximation, MQME-D combined with TSS offers an efficient way to obtain semi-quantitative decompositions of the dissipation in wide range of subsystem-bath couplings and temperatures. Even for the Brownian oscillator bath for which TSS does not significantly affect the BSD, TSS offered a useful way to improve the numerical convergence of MQME-D. However, TSS could not reproduce the asymmetry between the dissipation by two molecules, which becomes more pronounced toward the low-frequency region. We expect the asymmetry may be realized in our method by using separate scaling functions for different BSDs or extending our method to include non-equilibrium motion of the bath modes.\cite{Jang2002}

MQME-D shows quadratic scaling of the computational cost with the size of the system, and its parallelization across multiple processors is also straightforward. Moreover, the cost of MQME-D does not depend on the characteristic frequency of the bath mode, in contrast of HEOM-D which shows rapid increase of the burden as $\omega$ decreases. We therefore expect the usefulness of MQME-D to grow with larger systems for which the computational costs of numerically exact methods become expensive due to their exponential scaling and challenges in parallelization.



We anticipate that applying the framework outlined in Paper I\cite{Kim2024} to a range of quantum master equations would lead us to corresponding dissipation theories in the near future. Upon rigorous validation as demonstrated in this paper, these theories can be integrated with realistic Hamiltonian models extracted by using state-of-the-art experimental and computational techniques. \cite{Lee2017,Mennucci2019,Connors2021,Kim2023,Gustin2023} We envisage that such efforts would lead us to deeper understandings on the quantum dynamics in a wide range of systems including photosynthetic complexes,\cite{Wilkins2015,Blau2018,CardosoRamos2019} artificial excitonic systems,\cite{Collini2016,Yang2021,Bialas2022} plasmonic systems,\cite{Hsu2017,Bai2021} and molecular and solid-state qubits.\cite{Gertler2021,Harrington2022,Chiesa2023}


\section*{Supplementary Material}
See the Supplementary Material for comprehensive discussion and analysis on the artificial drift in the cumulative dissipation density $\mc{E}(\omega, t)$, which demonstrate that $\mc{E}(\omega, t)$ from HEOM can be reliably used as a quantitative benchmark after calibrating the drift.

\begin{acknowledgments}
CWK was financially supported by the National Research Foundation of Korea (NRF) grant funded by the Ministry of Science and ICT (MSIT) of Korea (Grant Number: 2022R1F1A1074027, 2023M3K5A1094813, and RS-2023-00218219). IF is supported by the National Science Foundation under Grant No. CHE-2102386 and PHY-2310657.
\end{acknowledgments}

\section*{Data Availability}
The data and computer programs that support the findings of this study are available from the corresponding author upon reasonable request.

\appendix
\section{A Brief Introduction to HEOM-D Method}\label{section:HEOM_D}
In this section, we present a minimal explanation about the motivation and formulation of HEOM-D,\cite{Kim2022} a technique we used for calculating the dissipation into a specific bath component in HEOM simulations. In HEOM, the RDM of the subsystem $\sigma(t)$ is propagated by coupled equations of motion which connect $\sigma(t)$ to the hierarchy of ADMs. It was reported that ADMs encode the consequence of subsystem-bath interaction during the dynamics, and therefore one can extract the statistics related to the bath from the ADMs.\cite{Zhu2012,Kato2016,Ikeda2022} However, such methods could only handle the bath modes in the entire BSD collectively, and does not allow isolation of the information regarding a single bath mode. Moreover, the widely used approach which re-classifies the bath mode of interest as the subsystem\cite{Womick2011,Novoderezhkin2017,Bennett2018} is not allowed for HEOM. This is because the subtraction of a bath mode from a BSD converts its analytical quantum time correlation function to a form which cannot be handled by HEOM without drastically increasing the complexity of the calculation.\cite{Yan2021}

The HEOM-D method\cite{Kim2022} overcomes this challenge by introducing an extra bath mode (``probe mode'') that weakly couples to the subsystem through the same channel as the bath mode under examination (``target mode''). Under such a setting, the dynamical information regarding the target mode can be elucidated from that of the probe mode. To formulate the method, we first divide the full Hamiltonian [\eq{eq:H}] into contributions of the target mode and the rest,
\begin{equation}
    \hat{H} = \hat{H}_\ti{bt} + \hat{H}_\ti{rest},
\end{equation}
respectively, where $\hat{H}_\ti{bt}$ can be factorized as $\hat{s} \otimes \hat{b}$ with $\hat{s}$ and $\hat{b}$ representing the subsystem and bath part of $\hat{H}_\ti{bt}$, respectively. For example, if we want to examine the $k$-th bath mode in the spin-boson Hamiltonian [Eqs.~(\ref{eq:H_sub_sb})--(\ref{eq:H_int_sb})], $\hat{s}$ and $\hat{b}$ needs to be set as
\begin{equation}
    \hat{s} = \ket{+}\bra{+}-\ket{-}\bra{-}, \quad \hat{b} = \frac{\hat{p}_k^2}{2} + \frac{\omega_k^2}{2} (\hat{x}_k - d_k)^2.
\end{equation}
We now modify the system Hamiltonian according to
\begin{equation}
    \hat{H}' = \hat{H} + \hat{H}_\ti{bp},
\end{equation}
where we added a new Hamiltonian component $\hat{H}_\ti{bp} = \alpha \hat{H}_\ti{bt}$ for the probe mode. The scaling constant $\alpha$ is the ratio between the coupling strengths of the target and probe modes, whose value should be small enough to ensure that the dynamics under $\hat{H}'$ remains almost identical to that under $\hat{H}$. As $\hat{H}_\ti{bp}$ was not a part of the original BSD and therefore does not alter the structure of HEOM, it can now be freely included in the subsystem and monitored over time. At the start of the dynamics, the density matrix $\hat{\rho}_\ti{bp}(0)$ for the probe mode is constructed as the same thermal equilibrium associated with the target mode. Then, the initial density $\hat{\sigma}'(0)$ for the extended subsystem $\hat{H}_\ti{sub} + \hat{H}_\ti{bp}$ is set to be $\hat{\sigma}(0) \otimes \hat{\rho}_\ti{bp}$ and propagated by the same structure of HEOM used for the original subsystem RDM $\hat{\sigma}(t)$. Practically, $\hat{\rho}_\ti{bp}$ is implemented by using a finite number $n$ of bath quantum states which faithfully represent the thermal equilibrium. The value of $n$ should be chosen carefully so as not to excessively increase the computational burden, as the time spent for applying perturbative low-temperature correction and RDM propagation increases with the order of $O(n^6)$ and $O(n^3)$, respectively (\stn{section:efficiency}). After propagating $\hat{\sigma}'(t)$ for a certain amount of time, the time-dependent dissipation $\Delta E_\ti{bp}(t)$ induced by the probe mode can be calculated by
\begin{equation}
    \Delta E_\ti{bp}(t) = \tr \big[\hat{H}_\ti{bp} \big\{ \hat{\sigma}'(t) - \hat{\sigma}'(0) \big\} \big].
\end{equation}
In Ref.~\citenum{Kim2022}, we proved that $\Delta E_\ti{bp}(t)$ is related to the dissipation $\Delta E_\ti{bt}(t)$ induced by the target mode via
\begin{equation}\label{eq:diss_bathmode}
    \Delta E_\ti{bt}(t) = \lim_{\alpha \rightarrow 0} \frac{\Delta E_\ti{bp}(t)}{\alpha^2}.
\end{equation}
For harmonic oscillator bath, \eq{eq:diss_bathmode} can be converted to the dissipation density by combining it with the definition of the BSD [\eq{eq:BSD}],
\begin{equation}\label{eq:diss_bathmode_HR}
    \mc{E}(\omega_\ti{bt}, t) = \lim_{s_\ti{bp} \rightarrow 0} \frac{J(\omega_\ti{bp})}{\hbar \omega_\ti{bp}^2 s_\ti{bp}} \Delta E_\ti{bp}(t),
\end{equation}
where $\omega_\ti{bt} = \omega_\ti{bp}$ is the frequency of the target and probe modes, and $s_\ti{bp}$ is the H-R factor of the probe mode
\begin{equation}\label{eq:HR_probe}
    s_\ti{bp} = \frac{\omega_\ti{bp} d_\ti{bp}^2}{2 \hbar}.
\end{equation}
with the corresponding PES displacement $d_\ti{bp}$. \Eq{eq:diss_bathmode_HR} states that $s_\ti{bp}$ must be sufficiently small to mimic the $s_\ti{bp} \rightarrow 0$ limit. In practice, however, reducing the value of $s_\ti{bp}$ too much negatively affects the accuracy of the calculation due to the limited machine precision. Therefore it is required to seek the balance between these two aspects by checking the convergence of the calculation with different values of $s_\ti{bp}$.

As a final remark, we note that the applicability of the approach illustrated in this Appendix is not limited to HEOM and is compatible with any numerically exact simulation methods for open quantum system dynamics,\cite{Diosi1998,Yan2014,Tamascelli2019,Varvelo2021,Cygorek2022,Kundu2023} whenever it is not possible to construct the extended subsystem by directly using the target mode.

\section{Discretization of the Bath Spectral Densities}\label{section:BSD_disc}
In this Appendix, we elaborate on how the BSDs are discretized for the MQME and MQME-D simulations. For the Drude-Lorentz spectral density [\eq{eq:DL_BSD}], we follow  Ref.~\citenum{Wang1999}, where the individual bath modes are placed at the frequencies
\begin{equation}\label{eq:DL_discrete}
    \omega_j = \frac{j^2}{N^2} \omega_\ti{max}, \quad j = 1, \: 2, \: \cdots, \: N,
\end{equation}
with $N$ being the total number of bath modes in a spectral density, and $\omega_\ti{max}$ the upper limit of the frequency. \Eq{eq:DL_discrete} makes the bath modes more densely packed in the low-frequency region to reflect the increase of the reorganization energy density $J_\ti{DL}(\omega) / \omega$ therein. If we now define a function $f_\ti{DL}(\omega)$ which connects the discretized [\eq{eq:BSD}] and continuous [\eq{eq:DL_BSD}] forms of the BSD via
\begin{equation}\label{eq:spd_rho_conn}
    \frac{\omega_j^3 d_j^2}{2} = \frac{J_\ti{DL}(\omega_j)}{f_\ti{DL}(\omega_j)},
\end{equation}
its explicit expression becomes
\begin{equation}\label{eq:rho_DL}
    f_\ti{DL}(\omega) = \frac{N}{2 \sqrt{\omega \omega_\ti{max}}},
\end{equation}
which renders the reorganization energies of the individual bath modes as
\begin{equation}\label{eq:reorg_disc_DL}
    \lambda_j = \frac{\omega_j^2 d_j^2}{2} = \frac{4 \Lambda}{j \pi} \frac{\omega_\ti{c} \omega_j}{\omega_j^2 + \omega_\ti{c}^2}.
\end{equation}
Meanwhile, the reorganization energy arising from the corresponding region of the continuous spectral density is
\begin{equation}\label{eq:reorg_cont_DL}
    \begin{split}
    \int_{(\omega_{j-1}+\omega_j)/2}^{(\omega_j+\omega_{j+1})/2} \frac{J_\ti{DL}(\omega)}{\omega} \: d \omega &\approx \frac{J_\ti{DL}(\omega_j)}{\omega_j} \bigg[ \bigg( \frac{\omega_j + \omega_{j+1}}{2} \bigg) - \bigg( \frac{\omega_{j-1} + \omega_j}{2} \bigg) \bigg] \\
    &= \frac{4 \Lambda}{j \pi} \frac{\omega_\ti{c} \omega_j}{\omega_j^2 + \omega_\ti{c}^2},
    \end{split}
\end{equation}
where the approximation becomes exact at the limit of $N \rightarrow \infty$. The equality between the last expressions of \eqs{eq:reorg_disc_DL}{eq:reorg_cont_DL} indicates the validity of $f_\ti{DL}(\omega)$ defined by \eq{eq:rho_DL}.

For the Brownian oscillator [\eq{eq:BO_BSD}], we set $\omega_0 < \omega_\ti{max}$ and calculate the frequency $\Omega$ where the reorganization energy density $J_\ti{BO} / \omega$ is maximized within $[0, \omega_\ti{max}]$, which is
\begin{equation}
    \Omega = \sqrt{\ti{max}[0, \omega_0^2 - 2 \gamma^2]}.
\end{equation}
If $\Omega = 0$, we can apply the same scheme as the Drude-Lorentz BSD [Eqs.~(\ref{eq:DL_discrete})--(\ref{eq:rho_DL})] with $J_\ti{DL}(\omega)$ in \eq{eq:spd_rho_conn} replaced by $J_\ti{BO}(\omega)$. Otherwise, we divide the frequency domain into two separate windows $[0, \Omega]$ and $(\Omega, \omega_\ti{max}]$ and describe each region by using half of the bath modes under separate discretization schemes. This is achieved by setting the bath frequencies $\{\omega_{1,j}\}$ ($\{\omega_{2,j}\}$) and the connecting function $f_\ti{BO1}(\omega)$ [$f_\ti{BO2}(\omega)$] for the former (latter) window as
\begin{subequations}\label{eq:BO_disc_1}
\begin{equation}
    \omega_{1,j} = \bigg[ 1 - \bigg( 1 - \frac{2j}{N} \bigg)^2 \bigg] \Omega, \quad f_\ti{BO1}(\omega) = \frac{N}{\sqrt{(\Omega - \omega) \Omega}}, \quad j = 1, \: 2, \: \cdots , \: \frac{N}{2} - 1,
\end{equation}
\begin{equation}
    \omega_{2,j} = \Omega + \frac{4j^2}{N^2} (\omega_\ti{max} - \Omega), \quad f_\ti{BO2}(\omega) = \frac{N}{\sqrt{(\omega - \Omega)(\omega_\ti{max} - \Omega)}}, \quad j = 1, \: 2, \: \cdots , \: \frac{N}{2}.
\end{equation}
\end{subequations}
\Eq{eq:BO_disc_1} lacks the case of $\omega = \Omega$ where both $f_\ti{BO1}(\omega)$ and $f_\ti{BO2}(\omega)$ diverge. Nevertheless, we can solve this issue by setting the reorganization energy of the discrete bath mode at $\omega = \Omega$ equal to that calculated from the continuous BSD $J_\ti{BO}(\omega)$ [\eq{eq:BO_BSD}] in the frequency window $[(\Omega - \omega_{1, N/2-1})/2, \: (\omega_{2, 1} - \Omega)/2]$. As a result, we get
\begin{equation}\label{eq:BO_disc_2}
    \lambda_{\omega = \Omega} = \frac{2 \Lambda}{\pi N^2} \frac{\omega_\ti{max} \omega_0^2}{\gamma (\omega_0^2 - \gamma^2)}.
\end{equation}
The discretization scheme defined by \eqs{eq:BO_disc_1}{eq:BO_disc_2} makes the bath modes more concentrated around $\omega = \Omega$ compared to the rest of the frequency domain. As a result, more emphasis is put on the region which contributes larger to the overall subsystem-bath coupling, as we did for $J_\ti{DL}(\omega)$.

\providecommand{\noopsort}[1]{}\providecommand{\singleletter}[1]{#1}%


\begin{thebibliography}{56}%
	\makeatletter
	\providecommand \@ifxundefined [1]{%
		\@ifx{#1\undefined}
	}%
	\providecommand \@ifnum [1]{%
		\ifnum #1\expandafter \@firstoftwo
		\else \expandafter \@secondoftwo
		\fi
	}%
	\providecommand \@ifx [1]{%
		\ifx #1\expandafter \@firstoftwo
		\else \expandafter \@secondoftwo
		\fi
	}%
	\providecommand \natexlab [1]{#1}%
	\providecommand \enquote  [1]{``#1''}%
	\providecommand \bibnamefont  [1]{#1}%
	\providecommand \bibfnamefont [1]{#1}%
	\providecommand \citenamefont [1]{#1}%
	\providecommand \href@noop [0]{\@secondoftwo}%
	\providecommand \href [0]{\begingroup \@sanitize@url \@href}%
	\providecommand \@href[1]{\@@startlink{#1}\@@href}%
	\providecommand \@@href[1]{\endgroup#1\@@endlink}%
	\providecommand \@sanitize@url [0]{\catcode `\\12\catcode `\$12\catcode `\&12\catcode `\#12\catcode `\^12\catcode `\_12\catcode `\%12\relax}%
	\providecommand \@@startlink[1]{}%
	\providecommand \@@endlink[0]{}%
	\providecommand \url  [0]{\begingroup\@sanitize@url \@url }%
	\providecommand \@url [1]{\endgroup\@href {#1}{\urlprefix }}%
	\providecommand \urlprefix  [0]{URL }%
	\providecommand \Eprint [0]{\href }%
	\providecommand \doibase [0]{https://doi.org/}%
	\providecommand \selectlanguage [0]{\@gobble}%
	\providecommand \bibinfo  [0]{\@secondoftwo}%
	\providecommand \bibfield  [0]{\@secondoftwo}%
	\providecommand \translation [1]{[#1]}%
	\providecommand \BibitemOpen [0]{}%
	\providecommand \bibitemStop [0]{}%
	\providecommand \bibitemNoStop [0]{.\EOS\space}%
	\providecommand \EOS [0]{\spacefactor3000\relax}%
	\providecommand \BibitemShut  [1]{\csname bibitem#1\endcsname}%
	\let\auto@bib@innerbib\@empty
	\bibitem [{\citenamefont {Kim}\ and\ \citenamefont {Franco}(2024)}]{Kim2024}%
	\BibitemOpen
	\bibfield  {author} {\bibinfo {author} {\bibfnamefont {C.~W.}\ \bibnamefont {Kim}}\ and\ \bibinfo {author} {\bibfnamefont {I.}~\bibnamefont {Franco}},\ }\bibfield  {title} {\enquote {\bibinfo {title} {{General Framework for Quantifying Dissipation Pathways in Open Quantum Systems. I. Theoretical Formulation}},}\ }\href@noop {} {\bibfield  {journal} {\bibinfo  {journal} {J. Chem. Phys.}\ }\textbf {\bibinfo {volume} {160}},\ \bibinfo {pages} {214111} (\bibinfo {year} {2024})}\BibitemShut {NoStop}%
	\bibitem [{\citenamefont {Nakajima}(1958)}]{Nakajima1958}%
	\BibitemOpen
	\bibfield  {author} {\bibinfo {author} {\bibfnamefont {S.}~\bibnamefont {Nakajima}},\ }\bibfield  {title} {\enquote {\bibinfo {title} {{On Quantum Theory of Transport Phenomena: Steady Diffusion}},}\ }\href {https://doi.org/10.1143/PTP.20.948} {\bibfield  {journal} {\bibinfo  {journal} {Prog. Theor. Phys.}\ }\textbf {\bibinfo {volume} {20}},\ \bibinfo {pages} {948--959} (\bibinfo {year} {1958})}\BibitemShut {NoStop}%
	\bibitem [{\citenamefont {Zwanzig}(1960)}]{Zwanzig1960}%
	\BibitemOpen
	\bibfield  {author} {\bibinfo {author} {\bibfnamefont {R.}~\bibnamefont {Zwanzig}},\ }\bibfield  {title} {\enquote {\bibinfo {title} {{Ensemble Method in the Theory of Irreversibility}},}\ }\href {https://doi.org/10.1063/1.1731409} {\bibfield  {journal} {\bibinfo  {journal} {J. Chem. Phys.}\ }\textbf {\bibinfo {volume} {33}},\ \bibinfo {pages} {1338--1341} (\bibinfo {year} {1960})}\BibitemShut {NoStop}%
	\bibitem [{\citenamefont {Kim}\ and\ \citenamefont {Franco}(2021)}]{Kim2021}%
	\BibitemOpen
	\bibfield  {author} {\bibinfo {author} {\bibfnamefont {C.~W.}\ \bibnamefont {Kim}}\ and\ \bibinfo {author} {\bibfnamefont {I.}~\bibnamefont {Franco}},\ }\bibfield  {title} {\enquote {\bibinfo {title} {{Theory of Dissipation Pathways in Open Quantum Systems}},}\ }\href {https://doi.org/10.1063/5.0038967} {\bibfield  {journal} {\bibinfo  {journal} {J. Chem. Phys.}\ }\textbf {\bibinfo {volume} {154}},\ \bibinfo {pages} {084109} (\bibinfo {year} {2021})}\BibitemShut {NoStop}%
	\bibitem [{\citenamefont {Kim}\ and\ \citenamefont {Rhee}(2020)}]{Kim2020}%
	\BibitemOpen
	\bibfield  {author} {\bibinfo {author} {\bibfnamefont {C.~W.}\ \bibnamefont {Kim}}\ and\ \bibinfo {author} {\bibfnamefont {Y.~M.}\ \bibnamefont {Rhee}},\ }\bibfield  {title} {\enquote {\bibinfo {title} {{Toward Monitoring the Dissipative Vibrational Energy Flows in Open Quantum Systems by Mixed Quantum-Classical Simulations}},}\ }\href {https://doi.org/10.1063/5.0009867} {\bibfield  {journal} {\bibinfo  {journal} {J. Chem. Phys.}\ }\textbf {\bibinfo {volume} {152}},\ \bibinfo {pages} {244109} (\bibinfo {year} {2020})}\BibitemShut {NoStop}%
	\bibitem [{\citenamefont {Nassimi}, \citenamefont {Bonella},\ and\ \citenamefont {Kapral}(2010)}]{Nassimi2010}%
	\BibitemOpen
	\bibfield  {author} {\bibinfo {author} {\bibfnamefont {A.}~\bibnamefont {Nassimi}}, \bibinfo {author} {\bibfnamefont {S.}~\bibnamefont {Bonella}},\ and\ \bibinfo {author} {\bibfnamefont {R.}~\bibnamefont {Kapral}},\ }\bibfield  {title} {\enquote {\bibinfo {title} {{Analysis of the Quantum-Classical Liouville Equation in the Mapping Basis}},}\ }\href {https://doi.org/10.1063/1.3480018} {\bibfield  {journal} {\bibinfo  {journal} {J. Chem. Phys.}\ }\textbf {\bibinfo {volume} {133}},\ \bibinfo {pages} {134115} (\bibinfo {year} {2010})}\BibitemShut {NoStop}%
	\bibitem [{\citenamefont {Kim}\ and\ \citenamefont {Rhee}(2014)}]{Kim2014}%
	\BibitemOpen
	\bibfield  {author} {\bibinfo {author} {\bibfnamefont {H.~W.}\ \bibnamefont {Kim}}\ and\ \bibinfo {author} {\bibfnamefont {Y.~M.}\ \bibnamefont {Rhee}},\ }\bibfield  {title} {\enquote {\bibinfo {title} {{Improving Long Time Behavior of Poisson Bracket Mapping Equation: A Non-Hamiltonian Approach}},}\ }\href {https://doi.org/10.1063/1.4874268} {\bibfield  {journal} {\bibinfo  {journal} {J. Chem. Phys.}\ }\textbf {\bibinfo {volume} {140}},\ \bibinfo {pages} {184106} (\bibinfo {year} {2014})}\BibitemShut {NoStop}%
	\bibitem [{\citenamefont {Kim}(2022)}]{Kim2022}%
	\BibitemOpen
	\bibfield  {author} {\bibinfo {author} {\bibfnamefont {C.~W.}\ \bibnamefont {Kim}},\ }\bibfield  {title} {\enquote {\bibinfo {title} {{Extracting Bath Information from Open-Quantum-System Dynamics with the Hierarchical Equations-of-Motion Method}},}\ }\href {https://doi.org/10.1103/PhysRevA.106.042223} {\bibfield  {journal} {\bibinfo  {journal} {Phys. Rev. A}\ }\textbf {\bibinfo {volume} {106}},\ \bibinfo {pages} {042223} (\bibinfo {year} {2022})}\BibitemShut {NoStop}%
	\bibitem [{\citenamefont {Tanimura}(1990)}]{Tanimura1990}%
	\BibitemOpen
	\bibfield  {author} {\bibinfo {author} {\bibfnamefont {Y.}~\bibnamefont {Tanimura}},\ }\bibfield  {title} {\enquote {\bibinfo {title} {{Nonperturbative Expansion Method for a Quantum System Coupled to a Harmonic-Oscillator Bath}},}\ }\href {https://doi.org/10.1103/PhysRevA.41.6676} {\bibfield  {journal} {\bibinfo  {journal} {Phys. Rev. A}\ }\textbf {\bibinfo {volume} {41}},\ \bibinfo {pages} {6676--6687} (\bibinfo {year} {1990})}\BibitemShut {NoStop}%
	\bibitem [{\citenamefont {Tanimura}(2020)}]{Tanimura2020}%
	\BibitemOpen
	\bibfield  {author} {\bibinfo {author} {\bibfnamefont {Y.}~\bibnamefont {Tanimura}},\ }\bibfield  {title} {\enquote {\bibinfo {title} {{Numerically “Exact” Approach to Open Quantum Dynamics: The Hierarchical Equations of Motion (HEOM)}},}\ }\href {https://doi.org/10.1063/5.0011599} {\bibfield  {journal} {\bibinfo  {journal} {J. Chem. Phys.}\ }\textbf {\bibinfo {volume} {153}},\ \bibinfo {pages} {020901} (\bibinfo {year} {2020})}\BibitemShut {NoStop}%
	\bibitem [{\citenamefont {Topaler}\ and\ \citenamefont {Makri}(1993)}]{Topaler1993}%
	\BibitemOpen
	\bibfield  {author} {\bibinfo {author} {\bibfnamefont {M.}~\bibnamefont {Topaler}}\ and\ \bibinfo {author} {\bibfnamefont {N.}~\bibnamefont {Makri}},\ }\bibfield  {title} {\enquote {\bibinfo {title} {{Quasi-Adiabatic Propagator Path Integral Methods. Exact Quantum Rate Constants for Condensed Phase Reactions}},}\ }\href {https://doi.org/https://doi.org/10.1016/0009-2614(93)89135-5} {\bibfield  {journal} {\bibinfo  {journal} {Chem. Phys. Lett.}\ }\textbf {\bibinfo {volume} {210}},\ \bibinfo {pages} {285--293} (\bibinfo {year} {1993})}\BibitemShut {NoStop}%
	\bibitem [{\citenamefont {Kundu}\ and\ \citenamefont {Makri}(2023)}]{Kundu2023}%
	\BibitemOpen
	\bibfield  {author} {\bibinfo {author} {\bibfnamefont {S.}~\bibnamefont {Kundu}}\ and\ \bibinfo {author} {\bibfnamefont {N.}~\bibnamefont {Makri}},\ }\bibfield  {title} {\enquote {\bibinfo {title} {{PathSum: A C++ and Fortran suite of fully quantum mechanical real-time path integral methods for (multi-)system + bath dynamics}},}\ }\href {https://doi.org/10.1063/5.0151748} {\bibfield  {journal} {\bibinfo  {journal} {J. Chem. Phys.}\ }\textbf {\bibinfo {volume} {158}},\ \bibinfo {pages} {224801} (\bibinfo {year} {2023})}\BibitemShut {NoStop}%
	\bibitem [{\citenamefont {Montoya-Castillo}, \citenamefont {Berkelbach},\ and\ \citenamefont {Reichman}(2015)}]{MontoyaCastillo2015}%
	\BibitemOpen
	\bibfield  {author} {\bibinfo {author} {\bibfnamefont {A.}~\bibnamefont {Montoya-Castillo}}, \bibinfo {author} {\bibfnamefont {T.~C.}\ \bibnamefont {Berkelbach}},\ and\ \bibinfo {author} {\bibfnamefont {D.~R.}\ \bibnamefont {Reichman}},\ }\bibfield  {title} {\enquote {\bibinfo {title} {{Extending the Applicability of Redfield Theories into Highly Non-Markovian Regimes}},}\ }\href {https://doi.org/10.1063/1.4935443} {\bibfield  {journal} {\bibinfo  {journal} {J. Chem. Phys.}\ }\textbf {\bibinfo {volume} {143}},\ \bibinfo {pages} {194108} (\bibinfo {year} {2015})}\BibitemShut {NoStop}%
	\bibitem [{\citenamefont {Ikeda}\ and\ \citenamefont {Scholes}(2020)}]{Ikeda2020}%
	\BibitemOpen
	\bibfield  {author} {\bibinfo {author} {\bibfnamefont {T.}~\bibnamefont {Ikeda}}\ and\ \bibinfo {author} {\bibfnamefont {G.~D.}\ \bibnamefont {Scholes}},\ }\bibfield  {title} {\enquote {\bibinfo {title} {Generalization of the hierarchical equations of motion theory for efficient calculations with arbitrary correlation functions},}\ }\href {https://doi.org/10.1063/5.0007327} {\bibfield  {journal} {\bibinfo  {journal} {J. Chem. Phys.}\ }\textbf {\bibinfo {volume} {152}},\ \bibinfo {pages} {204101} (\bibinfo {year} {2020})}\BibitemShut {NoStop}%
	\bibitem [{\citenamefont {O'Reilly}\ and\ \citenamefont {Olaya-Castro}(2014)}]{O'Reilly2014}%
	\BibitemOpen
	\bibfield  {author} {\bibinfo {author} {\bibfnamefont {E.~J.}\ \bibnamefont {O'Reilly}}\ and\ \bibinfo {author} {\bibfnamefont {A.}~\bibnamefont {Olaya-Castro}},\ }\bibfield  {title} {\enquote {\bibinfo {title} {{Non-Classicality of the Molecular Vibrations Assisting Exciton Energy Transfer at Room Temperature}},}\ }\href {https://doi.org/10.1038/ncomms4012} {\bibfield  {journal} {\bibinfo  {journal} {Nat. Commun.}\ }\textbf {\bibinfo {volume} {5}},\ \bibinfo {pages} {3012} (\bibinfo {year} {2014})}\BibitemShut {NoStop}%
	\bibitem [{\citenamefont {Novoderezhkin}\ \emph {et~al.}(2017)\citenamefont {Novoderezhkin}, \citenamefont {Romero}, \citenamefont {Prior},\ and\ \citenamefont {van Grondelle}}]{Novoderezhkin2017}%
	\BibitemOpen
	\bibfield  {author} {\bibinfo {author} {\bibfnamefont {V.~I.}\ \bibnamefont {Novoderezhkin}}, \bibinfo {author} {\bibfnamefont {E.}~\bibnamefont {Romero}}, \bibinfo {author} {\bibfnamefont {J.}~\bibnamefont {Prior}},\ and\ \bibinfo {author} {\bibfnamefont {R.}~\bibnamefont {van Grondelle}},\ }\bibfield  {title} {\enquote {\bibinfo {title} {Exciton-vibrational resonance and dynamics of charge separation in the photosystem ii reaction center},}\ }\href {http://dx.doi.org/10.1039/C6CP07308E} {\bibfield  {journal} {\bibinfo  {journal} {Phys. Chem. Chem. Phys.}\ }\textbf {\bibinfo {volume} {19}},\ \bibinfo {pages} {5195--5208} (\bibinfo {year} {2017})}\BibitemShut {NoStop}%
	\bibitem [{\citenamefont {Bennett}\ \emph {et~al.}(2018)\citenamefont {Bennett}, \citenamefont {Malý}, \citenamefont {Kreisbeck}, \citenamefont {van Grondelle},\ and\ \citenamefont {Aspuru-Guzik}}]{Bennett2018}%
	\BibitemOpen
	\bibfield  {author} {\bibinfo {author} {\bibfnamefont {D.~I.~G.}\ \bibnamefont {Bennett}}, \bibinfo {author} {\bibfnamefont {P.}~\bibnamefont {Malý}}, \bibinfo {author} {\bibfnamefont {C.}~\bibnamefont {Kreisbeck}}, \bibinfo {author} {\bibfnamefont {R.}~\bibnamefont {van Grondelle}},\ and\ \bibinfo {author} {\bibfnamefont {A.}~\bibnamefont {Aspuru-Guzik}},\ }\bibfield  {title} {\enquote {\bibinfo {title} {{Mechanistic Regimes of Vibronic Transport in a Heterodimer and the Design Principle of Incoherent Vibronic Transport in Phycobiliproteins}},}\ }\href {https://doi.org/10.1021/acs.jpclett.8b00844} {\bibfield  {journal} {\bibinfo  {journal} {J. Phys. Chem. Lett.}\ }\textbf {\bibinfo {volume} {9}},\ \bibinfo {pages} {2665--2670} (\bibinfo {year} {2018})}\BibitemShut {NoStop}%
	\bibitem [{\citenamefont {F{\"o}rster}(1959)}]{Forster1959}%
	\BibitemOpen
	\bibfield  {author} {\bibinfo {author} {\bibfnamefont {T.}~\bibnamefont {F{\"o}rster}},\ }\bibfield  {title} {\enquote {\bibinfo {title} {{10th Spiers Memorial Lecture. Transfer mechanisms of electronic excitation}},}\ }\href@noop {} {\bibfield  {journal} {\bibinfo  {journal} {Discuss. Faraday Soc.}\ }\textbf {\bibinfo {volume} {27}},\ \bibinfo {pages} {7--17} (\bibinfo {year} {1959})}\BibitemShut {NoStop}%
	\bibitem [{\citenamefont {Wang}\ \emph {et~al.}(1999)\citenamefont {Wang}, \citenamefont {Song}, \citenamefont {Chandler},\ and\ \citenamefont {Miller}}]{Wang1999}%
	\BibitemOpen
	\bibfield  {author} {\bibinfo {author} {\bibfnamefont {H.}~\bibnamefont {Wang}}, \bibinfo {author} {\bibfnamefont {X.}~\bibnamefont {Song}}, \bibinfo {author} {\bibfnamefont {D.}~\bibnamefont {Chandler}},\ and\ \bibinfo {author} {\bibfnamefont {W.~H.}\ \bibnamefont {Miller}},\ }\bibfield  {title} {\enquote {\bibinfo {title} {{Semiclassical Study of Electronically Nonadiabatic Dynamics in the Condensed-Phase: Spin-Boson Problem with Debye Spectral Sensity}},}\ }\href {https://doi.org/10.1063/1.478388} {\bibfield  {journal} {\bibinfo  {journal} {J. Chem. Phys.}\ }\textbf {\bibinfo {volume} {110}},\ \bibinfo {pages} {4828--4840} (\bibinfo {year} {1999})}\BibitemShut {NoStop}%
	\bibitem [{\citenamefont {Fay}(2022)}]{Fay2022}%
	\BibitemOpen
	\bibfield  {author} {\bibinfo {author} {\bibfnamefont {T.~P.}\ \bibnamefont {Fay}},\ }\bibfield  {title} {\enquote {\bibinfo {title} {{A Simple Improved Low Temperature Correction for the Hierarchical Equations of Motion}},}\ }\href {https://doi.org/10.1063/5.0100365} {\bibfield  {journal} {\bibinfo  {journal} {J. Chem. Phys.}\ }\textbf {\bibinfo {volume} {157}},\ \bibinfo {pages} {054108} (\bibinfo {year} {2022})}\BibitemShut {NoStop}%
	\bibitem [{\citenamefont {Fehlberg}(1969)}]{Fehlberg1969}%
	\BibitemOpen
	\bibfield  {author} {\bibinfo {author} {\bibfnamefont {E.}~\bibnamefont {Fehlberg}},\ }\href@noop {} {\emph {\bibinfo {title} {Low-order classical Runge-Kutta formulas with stepsize control and their application to some heat transfer problems}}},\ Vol.\ \bibinfo {volume} {315}\ (\bibinfo  {publisher} {National Aeronautics and Space Administration},\ \bibinfo {address} {Huntsville},\ \bibinfo {year} {1969})\BibitemShut {NoStop}%
	\bibitem [{\citenamefont {Tanimura}(2014)}]{Tanimura2014}%
	\BibitemOpen
	\bibfield  {author} {\bibinfo {author} {\bibfnamefont {Y.}~\bibnamefont {Tanimura}},\ }\bibfield  {title} {\enquote {\bibinfo {title} {{Reduced Hierarchical Equations of Motion in Real and Imaginary Time: Correlated Initial States and Thermodynamic Quantities}},}\ }\href {https://doi.org/10.1063/1.4890441} {\bibfield  {journal} {\bibinfo  {journal} {J. Chem. Phys.}\ }\textbf {\bibinfo {volume} {141}},\ \bibinfo {pages} {044114} (\bibinfo {year} {2014})}\BibitemShut {NoStop}%
	\bibitem [{\citenamefont {Berkelbach}, \citenamefont {Markland},\ and\ \citenamefont {Reichman}(2012)}]{Berkelbach2012}%
	\BibitemOpen
	\bibfield  {author} {\bibinfo {author} {\bibfnamefont {T.~C.}\ \bibnamefont {Berkelbach}}, \bibinfo {author} {\bibfnamefont {T.~E.}\ \bibnamefont {Markland}},\ and\ \bibinfo {author} {\bibfnamefont {D.~R.}\ \bibnamefont {Reichman}},\ }\bibfield  {title} {\enquote {\bibinfo {title} {{Reduced Density Matrix Hybrid Approach: Application to Electronic Energy Transfer}},}\ }\href {https://doi.org/10.1063/1.3687342} {\bibfield  {journal} {\bibinfo  {journal} {J. Chem. Phys.}\ }\textbf {\bibinfo {volume} {136}},\ \bibinfo {pages} {084104} (\bibinfo {year} {2012})}\BibitemShut {NoStop}%
	\bibitem [{\citenamefont {Ribeiro}\ \emph {et~al.}(2018)\citenamefont {Ribeiro}, \citenamefont {Mart{\'i}nez-Mart{\'i}nez}, \citenamefont {Du}, \citenamefont {Campos-Gonzalez-Angulo},\ and\ \citenamefont {Yuen-Zhou}}]{Ribeiro2018}%
	\BibitemOpen
	\bibfield  {author} {\bibinfo {author} {\bibfnamefont {R.~F.}\ \bibnamefont {Ribeiro}}, \bibinfo {author} {\bibfnamefont {L.~A.}\ \bibnamefont {Mart{\'i}nez-Mart{\'i}nez}}, \bibinfo {author} {\bibfnamefont {M.}~\bibnamefont {Du}}, \bibinfo {author} {\bibfnamefont {J.}~\bibnamefont {Campos-Gonzalez-Angulo}},\ and\ \bibinfo {author} {\bibfnamefont {J.}~\bibnamefont {Yuen-Zhou}},\ }\bibfield  {title} {\enquote {\bibinfo {title} {Polariton chemistry: controlling molecular dynamics with optical cavities},}\ }\href {https://doi.org/10.1039/C8SC01043A} {\bibfield  {journal} {\bibinfo  {journal} {Chem. Sci.}\ }\textbf {\bibinfo {volume} {9}},\ \bibinfo {pages} {6325--6339} (\bibinfo {year} {2018})}\BibitemShut {NoStop}%
	\bibitem [{\citenamefont {Ebbesen}, \citenamefont {Rubio},\ and\ \citenamefont {Scholes}(2023)}]{Ebbesen2023}%
	\BibitemOpen
	\bibfield  {author} {\bibinfo {author} {\bibfnamefont {T.~W.}\ \bibnamefont {Ebbesen}}, \bibinfo {author} {\bibfnamefont {A.}~\bibnamefont {Rubio}},\ and\ \bibinfo {author} {\bibfnamefont {G.~D.}\ \bibnamefont {Scholes}},\ }\bibfield  {title} {\enquote {\bibinfo {title} {Introduction: Polaritonic chemistry},}\ }\href {https://doi.org/10.1021/acs.chemrev.3c00637} {\bibfield  {journal} {\bibinfo  {journal} {Chem. Rev.}\ }\textbf {\bibinfo {volume} {123}},\ \bibinfo {pages} {12037--12038} (\bibinfo {year} {2023})}\BibitemShut {NoStop}%
	\bibitem [{\citenamefont {Varvelo}, \citenamefont {Lynd},\ and\ \citenamefont {Bennett}(2021)}]{Varvelo2021}%
	\BibitemOpen
	\bibfield  {author} {\bibinfo {author} {\bibfnamefont {L.}~\bibnamefont {Varvelo}}, \bibinfo {author} {\bibfnamefont {J.~K.}\ \bibnamefont {Lynd}},\ and\ \bibinfo {author} {\bibfnamefont {D.~I.~G.}\ \bibnamefont {Bennett}},\ }\bibfield  {title} {\enquote {\bibinfo {title} {Formally exact simulations of mesoscale exciton dynamics in molecular materials},}\ }\href {https://doi.org/10.1039/D1SC01448J} {\bibfield  {journal} {\bibinfo  {journal} {Chem. Sci.}\ }\textbf {\bibinfo {volume} {12}},\ \bibinfo {pages} {9704--9711} (\bibinfo {year} {2021})}\BibitemShut {NoStop}%
	\bibitem [{\citenamefont {Strümpfer}\ and\ \citenamefont {Schulten}(2012)}]{Johan2012}%
	\BibitemOpen
	\bibfield  {author} {\bibinfo {author} {\bibfnamefont {J.}~\bibnamefont {Strümpfer}}\ and\ \bibinfo {author} {\bibfnamefont {K.}~\bibnamefont {Schulten}},\ }\bibfield  {title} {\enquote {\bibinfo {title} {{Open Quantum Dynamics Calculations with the Hierarchy Equations of Motion on Parallel Computers}},}\ }\href {https://doi.org/10.1021/ct3003833} {\bibfield  {journal} {\bibinfo  {journal} {J. Chem. Theory Comput.}\ }\textbf {\bibinfo {volume} {8}},\ \bibinfo {pages} {2808--2816} (\bibinfo {year} {2012})}\BibitemShut {NoStop}%
	\bibitem [{\citenamefont {Noack}\ \emph {et~al.}(2018)\citenamefont {Noack}, \citenamefont {Reinefeld}, \citenamefont {Kramer},\ and\ \citenamefont {Steinke}}]{Noack2018}%
	\BibitemOpen
	\bibfield  {author} {\bibinfo {author} {\bibfnamefont {M.}~\bibnamefont {Noack}}, \bibinfo {author} {\bibfnamefont {A.}~\bibnamefont {Reinefeld}}, \bibinfo {author} {\bibfnamefont {T.}~\bibnamefont {Kramer}},\ and\ \bibinfo {author} {\bibfnamefont {T.}~\bibnamefont {Steinke}},\ }\bibfield  {title} {\enquote {\bibinfo {title} {{DM-HEOM: A Portable and Scalable Solver-Framework for the Hierarchical Equations of Motion}},}\ }in\ \href {https://doi.org/10.1109/IPDPSW.2018.00149} {\emph {\bibinfo {booktitle} {2018 IEEE International Parallel and Distributed Processing Symposium Workshops (IPDPSW)}}}\ (\bibinfo {year} {2018})\ pp.\ \bibinfo {pages} {947--956}\BibitemShut {NoStop}%
	\bibitem [{\citenamefont {Urbanek}\ and\ \citenamefont {Soldán}(2016)}]{Urbanek2016}%
	\BibitemOpen
	\bibfield  {author} {\bibinfo {author} {\bibfnamefont {M.}~\bibnamefont {Urbanek}}\ and\ \bibinfo {author} {\bibfnamefont {P.}~\bibnamefont {Soldán}},\ }\bibfield  {title} {\enquote {\bibinfo {title} {{Parallel implementation of the time-evolving block decimation algorithm for the Bose–Hubbard model}},}\ }\href {https://doi.org/https://doi.org/10.1016/j.cpc.2015.10.016} {\bibfield  {journal} {\bibinfo  {journal} {Comput. Phys. Commun.}\ }\textbf {\bibinfo {volume} {199}},\ \bibinfo {pages} {170--177} (\bibinfo {year} {2016})}\BibitemShut {NoStop}%
	\bibitem [{\citenamefont {Strathearn}\ \emph {et~al.}(2018)\citenamefont {Strathearn}, \citenamefont {Kirton}, \citenamefont {Kilda}, \citenamefont {Keeling},\ and\ \citenamefont {Lovett}}]{Strathearn2018}%
	\BibitemOpen
	\bibfield  {author} {\bibinfo {author} {\bibfnamefont {A.}~\bibnamefont {Strathearn}}, \bibinfo {author} {\bibfnamefont {P.}~\bibnamefont {Kirton}}, \bibinfo {author} {\bibfnamefont {D.}~\bibnamefont {Kilda}}, \bibinfo {author} {\bibfnamefont {J.}~\bibnamefont {Keeling}},\ and\ \bibinfo {author} {\bibfnamefont {B.~W.}\ \bibnamefont {Lovett}},\ }\bibfield  {title} {\enquote {\bibinfo {title} {Efficient non-markovian quantum dynamics using time-evolving matrix product operators},}\ }\href {https://doi.org/10.1038/s41467-018-05617-3} {\bibfield  {journal} {\bibinfo  {journal} {Nat. Commun.}\ }\textbf {\bibinfo {volume} {9}},\ \bibinfo {pages} {3322} (\bibinfo {year} {2018})}\BibitemShut {NoStop}%
	\bibitem [{\citenamefont {Jang}, \citenamefont {Jung},\ and\ \citenamefont {Silbey}(2002)}]{Jang2002}%
	\BibitemOpen
	\bibfield  {author} {\bibinfo {author} {\bibfnamefont {S.}~\bibnamefont {Jang}}, \bibinfo {author} {\bibfnamefont {Y.}~\bibnamefont {Jung}},\ and\ \bibinfo {author} {\bibfnamefont {R.~J.}\ \bibnamefont {Silbey}},\ }\bibfield  {title} {\enquote {\bibinfo {title} {{Nonequilibrium Generalization of Förster–Dexter Theory for Excitation Energy Transfer}},}\ }\href {https://doi.org/https://doi.org/10.1016/S0301-0104(01)00538-9} {\bibfield  {journal} {\bibinfo  {journal} {Chem. Phys.}\ }\textbf {\bibinfo {volume} {275}},\ \bibinfo {pages} {319--332} (\bibinfo {year} {2002})}\BibitemShut {NoStop}%
	\bibitem [{\citenamefont {Lee}, \citenamefont {Bravaya},\ and\ \citenamefont {Coker}(2017)}]{Lee2017}%
	\BibitemOpen
	\bibfield  {author} {\bibinfo {author} {\bibfnamefont {M.~K.}\ \bibnamefont {Lee}}, \bibinfo {author} {\bibfnamefont {K.~B.}\ \bibnamefont {Bravaya}},\ and\ \bibinfo {author} {\bibfnamefont {D.~F.}\ \bibnamefont {Coker}},\ }\bibfield  {title} {\enquote {\bibinfo {title} {{First-Principles Models for Biological Light-Harvesting: Phycobiliprotein Complexes from Cryptophyte Algae}},}\ }\href {https://doi.org/10.1021/jacs.7b01780} {\bibfield  {journal} {\bibinfo  {journal} {J. Am. Chem. Soc.}\ }\textbf {\bibinfo {volume} {139}},\ \bibinfo {pages} {7803--7814} (\bibinfo {year} {2017})}\BibitemShut {NoStop}%
	\bibitem [{\citenamefont {Mennucci}\ and\ \citenamefont {Corni}(2019)}]{Mennucci2019}%
	\BibitemOpen
	\bibfield  {author} {\bibinfo {author} {\bibfnamefont {B.}~\bibnamefont {Mennucci}}\ and\ \bibinfo {author} {\bibfnamefont {S.}~\bibnamefont {Corni}},\ }\bibfield  {title} {\enquote {\bibinfo {title} {{Multiscale Modelling of Photoinduced Processes in Composite Systems}},}\ }\href {https://doi.org/10.1038/s41570-019-0092-4} {\bibfield  {journal} {\bibinfo  {journal} {Nat. Rev. Chem.}\ }\textbf {\bibinfo {volume} {3}},\ \bibinfo {pages} {315--330} (\bibinfo {year} {2019})}\BibitemShut {NoStop}%
	\bibitem [{\citenamefont {Connors}\ \emph {et~al.}(2019)\citenamefont {Connors}, \citenamefont {Nelson}, \citenamefont {Qiao}, \citenamefont {Edge},\ and\ \citenamefont {Nichol}}]{Connors2021}%
	\BibitemOpen
	\bibfield  {author} {\bibinfo {author} {\bibfnamefont {E.~J.}\ \bibnamefont {Connors}}, \bibinfo {author} {\bibfnamefont {J.}~\bibnamefont {Nelson}}, \bibinfo {author} {\bibfnamefont {H.}~\bibnamefont {Qiao}}, \bibinfo {author} {\bibfnamefont {L.~F.}\ \bibnamefont {Edge}},\ and\ \bibinfo {author} {\bibfnamefont {J.~M.}\ \bibnamefont {Nichol}},\ }\bibfield  {title} {\enquote {\bibinfo {title} {{Low-frequency charge noise in Si/SiGe quantum dots}},}\ }\href {https://doi.org/10.1103/PhysRevB.100.165305} {\bibfield  {journal} {\bibinfo  {journal} {Phys. Rev. B}\ }\textbf {\bibinfo {volume} {100}},\ \bibinfo {pages} {165305} (\bibinfo {year} {2019})}\BibitemShut {NoStop}%
	\bibitem [{\citenamefont {Kim}\ \emph {et~al.}(2023)\citenamefont {Kim}, \citenamefont {Mitchell}, \citenamefont {Lawrence}, \citenamefont {Morozov}, \citenamefont {Savikhin},\ and\ \citenamefont {Slipchenko}}]{Kim2023}%
	\BibitemOpen
	\bibfield  {author} {\bibinfo {author} {\bibfnamefont {Y.}~\bibnamefont {Kim}}, \bibinfo {author} {\bibfnamefont {Z.}~\bibnamefont {Mitchell}}, \bibinfo {author} {\bibfnamefont {J.}~\bibnamefont {Lawrence}}, \bibinfo {author} {\bibfnamefont {D.}~\bibnamefont {Morozov}}, \bibinfo {author} {\bibfnamefont {S.}~\bibnamefont {Savikhin}},\ and\ \bibinfo {author} {\bibfnamefont {L.~V.}\ \bibnamefont {Slipchenko}},\ }\bibfield  {title} {\enquote {\bibinfo {title} {{Predicting Mutation-Induced Changes in the Electronic Properties of Photosynthetic Proteins from First Principles: The Fenna–Matthews–Olson Complex Example}},}\ }\href {https://doi.org/10.1021/acs.jpclett.3c01461} {\bibfield  {journal} {\bibinfo  {journal} {J. Phys. Chem. Lett.}\ }\textbf {\bibinfo {volume} {14}},\ \bibinfo {pages} {7038--7044} (\bibinfo {year} {2023})}\BibitemShut {NoStop}%
	\bibitem [{\citenamefont {Gustin}\ \emph {et~al.}(2023)\citenamefont {Gustin}, \citenamefont {Kim}, \citenamefont {McCamant},\ and\ \citenamefont {Franco}}]{Gustin2023}%
	\BibitemOpen
	\bibfield  {author} {\bibinfo {author} {\bibfnamefont {I.}~\bibnamefont {Gustin}}, \bibinfo {author} {\bibfnamefont {C.~W.}\ \bibnamefont {Kim}}, \bibinfo {author} {\bibfnamefont {D.~W.}\ \bibnamefont {McCamant}},\ and\ \bibinfo {author} {\bibfnamefont {I.}~\bibnamefont {Franco}},\ }\bibfield  {title} {\enquote {\bibinfo {title} {{Mapping Electronic Decoherence Pathways in Molecules}},}\ }\href@noop {} {\bibfield  {journal} {\bibinfo  {journal} {Proc. Natl. Acad. Sci. U. S. A.}\ }\textbf {\bibinfo {volume} {120}},\ \bibinfo {pages} {e2309987120} (\bibinfo {year} {2023})}\BibitemShut {NoStop}%
	\bibitem [{\citenamefont {Wilkins}\ and\ \citenamefont {Dattani}(2015)}]{Wilkins2015}%
	\BibitemOpen
	\bibfield  {author} {\bibinfo {author} {\bibfnamefont {D.~M.}\ \bibnamefont {Wilkins}}\ and\ \bibinfo {author} {\bibfnamefont {N.~S.}\ \bibnamefont {Dattani}},\ }\bibfield  {title} {\enquote {\bibinfo {title} {{Why Quantum Coherence Is Not Important in the Fenna–Matthews–Olsen Complex}},}\ }\href {https://doi.org/10.1021/ct501066k} {\bibfield  {journal} {\bibinfo  {journal} {J. Chem. Theory Comput.}\ }\textbf {\bibinfo {volume} {11}},\ \bibinfo {pages} {3411--3419} (\bibinfo {year} {2015})}\BibitemShut {NoStop}%
	\bibitem [{\citenamefont {Blau}\ \emph {et~al.}(2018)\citenamefont {Blau}, \citenamefont {Bennett}, \citenamefont {Kreisbeck}, \citenamefont {Scholes},\ and\ \citenamefont {Aspuru-Guzik}}]{Blau2018}%
	\BibitemOpen
	\bibfield  {author} {\bibinfo {author} {\bibfnamefont {S.~M.}\ \bibnamefont {Blau}}, \bibinfo {author} {\bibfnamefont {D.~I.~G.}\ \bibnamefont {Bennett}}, \bibinfo {author} {\bibfnamefont {C.}~\bibnamefont {Kreisbeck}}, \bibinfo {author} {\bibfnamefont {G.~D.}\ \bibnamefont {Scholes}},\ and\ \bibinfo {author} {\bibfnamefont {A.}~\bibnamefont {Aspuru-Guzik}},\ }\bibfield  {title} {\enquote {\bibinfo {title} {{Local Protein Solvation Drives Direct Down-Conversion in Phycobiliprotein PC645 via Incoherent Vibronic Transport}},}\ }\href {https://doi.org/10.1073/pnas.1800370115} {\bibfield  {journal} {\bibinfo  {journal} {Proc. Natl. Acad. Sci. U. S. A.}\ }\textbf {\bibinfo {volume} {115}},\ \bibinfo {pages} {E3342--E3350} (\bibinfo {year} {2018})}\BibitemShut {NoStop}%
	\bibitem [{\citenamefont {Cardoso~Ramos}\ \emph {et~al.}(2019)\citenamefont {Cardoso~Ramos}, \citenamefont {Nottoli}, \citenamefont {Cupellini},\ and\ \citenamefont {Mennucci}}]{CardosoRamos2019}%
	\BibitemOpen
	\bibfield  {author} {\bibinfo {author} {\bibfnamefont {F.}~\bibnamefont {Cardoso~Ramos}}, \bibinfo {author} {\bibfnamefont {M.}~\bibnamefont {Nottoli}}, \bibinfo {author} {\bibfnamefont {L.}~\bibnamefont {Cupellini}},\ and\ \bibinfo {author} {\bibfnamefont {B.}~\bibnamefont {Mennucci}},\ }\bibfield  {title} {\enquote {\bibinfo {title} {{The Molecular Mechanisms of Light Adaption in Light-Harvesting Complexes of Purple Bacteria Revealed by a Multiscale Modeling}},}\ }\href {https://doi.org/10.1039/C9SC02886B} {\bibfield  {journal} {\bibinfo  {journal} {Chem. Sci.}\ }\textbf {\bibinfo {volume} {10}},\ \bibinfo {pages} {9650--9662} (\bibinfo {year} {2019})}\BibitemShut {NoStop}%
	\bibitem [{\citenamefont {Bolzonello}, \citenamefont {Fassioli},\ and\ \citenamefont {Collini}(2016)}]{Collini2016}%
	\BibitemOpen
	\bibfield  {author} {\bibinfo {author} {\bibfnamefont {L.}~\bibnamefont {Bolzonello}}, \bibinfo {author} {\bibfnamefont {F.}~\bibnamefont {Fassioli}},\ and\ \bibinfo {author} {\bibfnamefont {E.}~\bibnamefont {Collini}},\ }\bibfield  {title} {\enquote {\bibinfo {title} {{Correlated Fluctuations and Intraband Dynamics of J-Aggregates Revealed by Combination of 2DES Schemes}},}\ }\href {https://doi.org/10.1021/acs.jpclett.6b02433} {\bibfield  {journal} {\bibinfo  {journal} {J. Phys. Chem. Lett.}\ }\textbf {\bibinfo {volume} {7}},\ \bibinfo {pages} {4996--5001} (\bibinfo {year} {2016})}\BibitemShut {NoStop}%
	\bibitem [{\citenamefont {Yang}\ and\ \citenamefont {Jang}(2020)}]{Yang2021}%
	\BibitemOpen
	\bibfield  {author} {\bibinfo {author} {\bibfnamefont {L.}~\bibnamefont {Yang}}\ and\ \bibinfo {author} {\bibfnamefont {S.~J.}\ \bibnamefont {Jang}},\ }\bibfield  {title} {\enquote {\bibinfo {title} {{Theoretical Investigation of Non-Förster Exciton Transfer Mechanisms in Perylene Diimide Donor, Phenylene Bridge, and Terrylene Diimide Acceptor Systems}},}\ }\href {https://doi.org/10.1063/5.0023709} {\bibfield  {journal} {\bibinfo  {journal} {J. Chem. Phys.}\ }\textbf {\bibinfo {volume} {153}},\ \bibinfo {pages} {144305} (\bibinfo {year} {2020})}\BibitemShut {NoStop}%
	\bibitem [{\citenamefont {Bialas}\ and\ \citenamefont {Spano}(2022)}]{Bialas2022}%
	\BibitemOpen
	\bibfield  {author} {\bibinfo {author} {\bibfnamefont {A.~L.}\ \bibnamefont {Bialas}}\ and\ \bibinfo {author} {\bibfnamefont {F.~C.}\ \bibnamefont {Spano}},\ }\bibfield  {title} {\enquote {\bibinfo {title} {{A Holstein–Peierls Approach to Excimer Spectra: The Evolution from Vibronically Structured to Unstructured Emission}},}\ }\href {https://doi.org/10.1021/acs.jpcc.1c10255} {\bibfield  {journal} {\bibinfo  {journal} {J. Phys. Chem. C}\ }\textbf {\bibinfo {volume} {126}},\ \bibinfo {pages} {4067--4081} (\bibinfo {year} {2022})}\BibitemShut {NoStop}%
	\bibitem [{\citenamefont {Hsu}, \citenamefont {Ding},\ and\ \citenamefont {Schatz}(2017)}]{Hsu2017}%
	\BibitemOpen
	\bibfield  {author} {\bibinfo {author} {\bibfnamefont {L.-Y.}\ \bibnamefont {Hsu}}, \bibinfo {author} {\bibfnamefont {W.}~\bibnamefont {Ding}},\ and\ \bibinfo {author} {\bibfnamefont {G.~C.}\ \bibnamefont {Schatz}},\ }\bibfield  {title} {\enquote {\bibinfo {title} {{Plasmon-Coupled Resonance Energy Transfer}},}\ }\href {https://doi.org/10.1021/acs.jpclett.7b00526} {\bibfield  {journal} {\bibinfo  {journal} {J. Phys. Chem. Lett.}\ }\textbf {\bibinfo {volume} {8}},\ \bibinfo {pages} {2357--2367} (\bibinfo {year} {2017})}\BibitemShut {NoStop}%
	\bibitem [{\citenamefont {Bai}\ \emph {et~al.}(2021)\citenamefont {Bai}, \citenamefont {ter Huurne}, \citenamefont {van Heijst}, \citenamefont {Murai},\ and\ \citenamefont {G{\'o}mez~Rivas}}]{Bai2021}%
	\BibitemOpen
	\bibfield  {author} {\bibinfo {author} {\bibfnamefont {P.}~\bibnamefont {Bai}}, \bibinfo {author} {\bibfnamefont {S.}~\bibnamefont {ter Huurne}}, \bibinfo {author} {\bibfnamefont {E.}~\bibnamefont {van Heijst}}, \bibinfo {author} {\bibfnamefont {S.}~\bibnamefont {Murai}},\ and\ \bibinfo {author} {\bibfnamefont {J.}~\bibnamefont {G{\'o}mez~Rivas}},\ }\bibfield  {title} {\enquote {\bibinfo {title} {{Evolutionary Optimization of Light-Matter Coupling in Open Plasmonic Cavities}},}\ }\href {https://doi.org/10.1063/5.0042056} {\bibfield  {journal} {\bibinfo  {journal} {J. Chem. Phys.}\ }\textbf {\bibinfo {volume} {154}},\ \bibinfo {pages} {134110} (\bibinfo {year} {2021})}\BibitemShut {NoStop}%
	\bibitem [{\citenamefont {Gertler}\ \emph {et~al.}(2021)\citenamefont {Gertler}, \citenamefont {Baker}, \citenamefont {Li}, \citenamefont {Shirol}, \citenamefont {Koch},\ and\ \citenamefont {Wang}}]{Gertler2021}%
	\BibitemOpen
	\bibfield  {author} {\bibinfo {author} {\bibfnamefont {J.~M.}\ \bibnamefont {Gertler}}, \bibinfo {author} {\bibfnamefont {B.}~\bibnamefont {Baker}}, \bibinfo {author} {\bibfnamefont {J.}~\bibnamefont {Li}}, \bibinfo {author} {\bibfnamefont {S.}~\bibnamefont {Shirol}}, \bibinfo {author} {\bibfnamefont {J.}~\bibnamefont {Koch}},\ and\ \bibinfo {author} {\bibfnamefont {C.}~\bibnamefont {Wang}},\ }\bibfield  {title} {\enquote {\bibinfo {title} {{Protecting a Bosonic Qubit with Autonomous Quantum Error Correction}},}\ }\href {https://doi.org/10.1038/s41586-021-03257-0} {\bibfield  {journal} {\bibinfo  {journal} {Nature}\ }\textbf {\bibinfo {volume} {590}},\ \bibinfo {pages} {243--248} (\bibinfo {year} {2021})}\BibitemShut {NoStop}%
	\bibitem [{\citenamefont {Harrington}, \citenamefont {Mueller},\ and\ \citenamefont {Murch}(2022)}]{Harrington2022}%
	\BibitemOpen
	\bibfield  {author} {\bibinfo {author} {\bibfnamefont {P.~M.}\ \bibnamefont {Harrington}}, \bibinfo {author} {\bibfnamefont {E.~J.}\ \bibnamefont {Mueller}},\ and\ \bibinfo {author} {\bibfnamefont {K.~W.}\ \bibnamefont {Murch}},\ }\bibfield  {title} {\enquote {\bibinfo {title} {{Engineered Dissipation for Quantum Information Science}},}\ }\href {https://doi.org/10.1038/s42254-022-00494-8} {\bibfield  {journal} {\bibinfo  {journal} {Nat. Rev. Phys.}\ }\textbf {\bibinfo {volume} {4}},\ \bibinfo {pages} {660--671} (\bibinfo {year} {2022})}\BibitemShut {NoStop}%
	\bibitem [{\citenamefont {Chiesa}\ \emph {et~al.}(2023)\citenamefont {Chiesa}, \citenamefont {Privitera}, \citenamefont {Macaluso}, \citenamefont {Mannini}, \citenamefont {Bittl}, \citenamefont {Naaman}, \citenamefont {Wasielewski}, \citenamefont {Sessoli},\ and\ \citenamefont {Carretta}}]{Chiesa2023}%
	\BibitemOpen
	\bibfield  {author} {\bibinfo {author} {\bibfnamefont {A.}~\bibnamefont {Chiesa}}, \bibinfo {author} {\bibfnamefont {A.}~\bibnamefont {Privitera}}, \bibinfo {author} {\bibfnamefont {E.}~\bibnamefont {Macaluso}}, \bibinfo {author} {\bibfnamefont {M.}~\bibnamefont {Mannini}}, \bibinfo {author} {\bibfnamefont {R.}~\bibnamefont {Bittl}}, \bibinfo {author} {\bibfnamefont {R.}~\bibnamefont {Naaman}}, \bibinfo {author} {\bibfnamefont {M.~R.}\ \bibnamefont {Wasielewski}}, \bibinfo {author} {\bibfnamefont {R.}~\bibnamefont {Sessoli}},\ and\ \bibinfo {author} {\bibfnamefont {S.}~\bibnamefont {Carretta}},\ }\bibfield  {title} {\enquote {\bibinfo {title} {Chirality-induced spin selectivity: An enabling technology for quantum applications},}\ }\href {https://doi.org/https://doi.org/10.1002/adma.202300472} {\bibfield  {journal} {\bibinfo  {journal} {Adv. Mater.}\ }\textbf {\bibinfo {volume} {35}},\ \bibinfo {pages} {2300472} (\bibinfo {year} {2023})}\BibitemShut {NoStop}%
	\bibitem [{\citenamefont {Zhu}\ \emph {et~al.}(2012)\citenamefont {Zhu}, \citenamefont {Liu}, \citenamefont {Xie},\ and\ \citenamefont {Shi}}]{Zhu2012}%
	\BibitemOpen
	\bibfield  {author} {\bibinfo {author} {\bibfnamefont {L.}~\bibnamefont {Zhu}}, \bibinfo {author} {\bibfnamefont {H.}~\bibnamefont {Liu}}, \bibinfo {author} {\bibfnamefont {W.}~\bibnamefont {Xie}},\ and\ \bibinfo {author} {\bibfnamefont {Q.}~\bibnamefont {Shi}},\ }\bibfield  {title} {\enquote {\bibinfo {title} {{Explicit system-bath correlation calculated using the hierarchical equations of motion method}},}\ }\href {https://doi.org/10.1063/1.4766358} {\bibfield  {journal} {\bibinfo  {journal} {J. Chem. Phys.}\ }\textbf {\bibinfo {volume} {137}},\ \bibinfo {pages} {194106} (\bibinfo {year} {2012})}\BibitemShut {NoStop}%
	\bibitem [{\citenamefont {Kato}\ and\ \citenamefont {Tanimura}(2016)}]{Kato2016}%
	\BibitemOpen
	\bibfield  {author} {\bibinfo {author} {\bibfnamefont {A.}~\bibnamefont {Kato}}\ and\ \bibinfo {author} {\bibfnamefont {Y.}~\bibnamefont {Tanimura}},\ }\bibfield  {title} {\enquote {\bibinfo {title} {{Quantum heat current under non-perturbative and non-Markovian conditions: Applications to heat machines}},}\ }\href@noop {} {\bibfield  {journal} {\bibinfo  {journal} {J. Chem. Phys.}\ }\textbf {\bibinfo {volume} {145}},\ \bibinfo {pages} {224105} (\bibinfo {year} {2016})}\BibitemShut {NoStop}%
	\bibitem [{\citenamefont {Ikeda}\ and\ \citenamefont {Nakayama}(2022)}]{Ikeda2022}%
	\BibitemOpen
	\bibfield  {author} {\bibinfo {author} {\bibfnamefont {T.}~\bibnamefont {Ikeda}}\ and\ \bibinfo {author} {\bibfnamefont {A.}~\bibnamefont {Nakayama}},\ }\bibfield  {title} {\enquote {\bibinfo {title} {{Collective bath coordinate mapping of “hierarchy” in hierarchical equations of motion}},}\ }\href {https://doi.org/10.1063/5.0082936} {\bibfield  {journal} {\bibinfo  {journal} {J. Chem. Phys.}\ }\textbf {\bibinfo {volume} {156}},\ \bibinfo {pages} {104104} (\bibinfo {year} {2022})}\BibitemShut {NoStop}%
	\bibitem [{\citenamefont {Womick}\ and\ \citenamefont {Moran}(2011)}]{Womick2011}%
	\BibitemOpen
	\bibfield  {author} {\bibinfo {author} {\bibfnamefont {J.~M.}\ \bibnamefont {Womick}}\ and\ \bibinfo {author} {\bibfnamefont {A.~M.}\ \bibnamefont {Moran}},\ }\bibfield  {title} {\enquote {\bibinfo {title} {Vibronic enhancement of exciton sizes and energy transport in photosynthetic complexes},}\ }\href {https://doi.org/10.1021/jp106713q} {\bibfield  {journal} {\bibinfo  {journal} {J. Phys. Chem. B}\ }\textbf {\bibinfo {volume} {115}},\ \bibinfo {pages} {1347--1356} (\bibinfo {year} {2011})}\BibitemShut {NoStop}%
	\bibitem [{\citenamefont {Yan}\ \emph {et~al.}(2021)\citenamefont {Yan}, \citenamefont {Xu}, \citenamefont {Li},\ and\ \citenamefont {Shi}}]{Yan2021}%
	\BibitemOpen
	\bibfield  {author} {\bibinfo {author} {\bibfnamefont {Y.}~\bibnamefont {Yan}}, \bibinfo {author} {\bibfnamefont {M.}~\bibnamefont {Xu}}, \bibinfo {author} {\bibfnamefont {T.}~\bibnamefont {Li}},\ and\ \bibinfo {author} {\bibfnamefont {Q.}~\bibnamefont {Shi}},\ }\bibfield  {title} {\enquote {\bibinfo {title} {{Efficient propagation of the hierarchical equations of motion using the Tucker and hierarchical Tucker tensors}},}\ }\href {https://doi.org/10.1063/5.0050720} {\bibfield  {journal} {\bibinfo  {journal} {J. Chem. Phys.}\ }\textbf {\bibinfo {volume} {154}},\ \bibinfo {pages} {194104} (\bibinfo {year} {2021})}\BibitemShut {NoStop}%
	\bibitem [{\citenamefont {Di{\'o}si}, \citenamefont {Gisin},\ and\ \citenamefont {Strunz}(1998)}]{Diosi1998}%
	\BibitemOpen
	\bibfield  {author} {\bibinfo {author} {\bibfnamefont {L.}~\bibnamefont {Di{\'o}si}}, \bibinfo {author} {\bibfnamefont {N.}~\bibnamefont {Gisin}},\ and\ \bibinfo {author} {\bibfnamefont {W.~T.}\ \bibnamefont {Strunz}},\ }\bibfield  {title} {\enquote {\bibinfo {title} {Non-markovian quantum state diffusion},}\ }\href {https://doi.org/10.1103/PhysRevA.58.1699} {\bibfield  {journal} {\bibinfo  {journal} {Phys. Rev. A}\ }\textbf {\bibinfo {volume} {58}},\ \bibinfo {pages} {1699--1712} (\bibinfo {year} {1998})}\BibitemShut {NoStop}%
	\bibitem [{\citenamefont {Yan}(2014)}]{Yan2014}%
	\BibitemOpen
	\bibfield  {author} {\bibinfo {author} {\bibfnamefont {Y.}~\bibnamefont {Yan}},\ }\bibfield  {title} {\enquote {\bibinfo {title} {{Theory of open quantum systems with bath of electrons and phonons and spins: Many-dissipaton density matrixes approach}},}\ }\href {https://doi.org/10.1063/1.4863379} {\bibfield  {journal} {\bibinfo  {journal} {J. Chem. Phys.}\ }\textbf {\bibinfo {volume} {140}},\ \bibinfo {pages} {054105} (\bibinfo {year} {2014})}\BibitemShut {NoStop}%
	\bibitem [{\citenamefont {Tamascelli}\ \emph {et~al.}(2019)\citenamefont {Tamascelli}, \citenamefont {Smirne}, \citenamefont {Lim}, \citenamefont {Huelga},\ and\ \citenamefont {Plenio}}]{Tamascelli2019}%
	\BibitemOpen
	\bibfield  {author} {\bibinfo {author} {\bibfnamefont {D.}~\bibnamefont {Tamascelli}}, \bibinfo {author} {\bibfnamefont {A.}~\bibnamefont {Smirne}}, \bibinfo {author} {\bibfnamefont {J.}~\bibnamefont {Lim}}, \bibinfo {author} {\bibfnamefont {S.~F.}\ \bibnamefont {Huelga}},\ and\ \bibinfo {author} {\bibfnamefont {M.~B.}\ \bibnamefont {Plenio}},\ }\bibfield  {title} {\enquote {\bibinfo {title} {Efficient simulation of finite-temperature open quantum systems},}\ }\href {https://link.aps.org/doi/10.1103/PhysRevLett.123.090402} {\bibfield  {journal} {\bibinfo  {journal} {Phys. Rev. Lett.}\ }\textbf {\bibinfo {volume} {123}},\ \bibinfo {pages} {090402} (\bibinfo {year} {2019})}\BibitemShut {NoStop}%
	\bibitem [{\citenamefont {Cygorek}\ \emph {et~al.}(2022)\citenamefont {Cygorek}, \citenamefont {Cosacchi}, \citenamefont {Vagov}, \citenamefont {Axt}, \citenamefont {Lovett}, \citenamefont {Keeling},\ and\ \citenamefont {Gauger}}]{Cygorek2022}%
	\BibitemOpen
	\bibfield  {author} {\bibinfo {author} {\bibfnamefont {M.}~\bibnamefont {Cygorek}}, \bibinfo {author} {\bibfnamefont {M.}~\bibnamefont {Cosacchi}}, \bibinfo {author} {\bibfnamefont {A.}~\bibnamefont {Vagov}}, \bibinfo {author} {\bibfnamefont {V.~M.}\ \bibnamefont {Axt}}, \bibinfo {author} {\bibfnamefont {B.~W.}\ \bibnamefont {Lovett}}, \bibinfo {author} {\bibfnamefont {J.}~\bibnamefont {Keeling}},\ and\ \bibinfo {author} {\bibfnamefont {E.~M.}\ \bibnamefont {Gauger}},\ }\bibfield  {title} {\enquote {\bibinfo {title} {Simulation of open quantum systems by automated compression of arbitrary environments},}\ }\href {https://doi.org/10.1038/s41567-022-01544-9} {\bibfield  {journal} {\bibinfo  {journal} {Nat. Phys.}\ }\textbf {\bibinfo {volume} {18}},\ \bibinfo {pages} {662--668} (\bibinfo {year} {2022})}\BibitemShut {NoStop}%
\end{thebibliography}
\end{document}